\documentclass[11pt]{article}

\usepackage{verbatim}
\usepackage{amsmath}
\usepackage{amsfonts}
\usepackage{amssymb}
\usepackage{bbm}
\usepackage{latexsym}
\usepackage{graphics}
\usepackage{lscape} 
\usepackage{epsfig}
\usepackage{color}

\usepackage{cite}

\renewcommand{\d}{\mathrm{d}}

\newcommand{\captn}[1]{\vspace{-3ex}\caption{\small #1}}

\DeclareMathSymbol{\mg}{\mathrel}{symbols}{"1D}

\newcommand{\gd}{\delta}
\renewcommand{\ge}{\epsilon}

\newcommand{\gf}{\phi}
\newcommand{\gvf}{\varphi}

\newcommand{\gx}{\xi}

\newcommand{\gth}{\theta}

\newcommand{\gp}{\pi}

\newcommand{\gch}{\chi}
\newcommand{\gTh}{\Theta}
\newcommand{\gO}{\Omega}

\newcommand{\cA}{{\cal A}}
\newcommand{\cB}{{\cal B}}

\newcommand{\cF}{{\cal F}}

\newcommand{\cH}{{\cal H}}

\newcommand{\cK}{{\cal K}}

\newcommand{\cM}{{\cal M}}

\newcommand{\cR}{{\cal R}}

\newcommand{\uA}{{\underline A}}
\newcommand{\uB}{{\underline B}}

\newcommand{\tB}{{\tilde B}}

\newcommand{\tZ}{{\tilde Z}}
\newcommand{\Tr}{\mbox{Tr}}
\newcommand{\tr}{\text{tr}}

\newcommand{\Id}{\text{\small 1}\hspace{-3.5pt}\text{1}}

\newcommand{\ra}{\rightarrow}

\newcommand{\der}{\partial}

\newcommand{\inv}{^{-1}}
\newcommand{\dsp}{\displaystyle}

\newcommand{\undr}[1]{{\underline{#1}}}

\newcommand{\labl}[1]{\label{#1}}
\newcommand{\half}{\frac 12 }
\newcommand{\shalf}{{\scriptstyle \half}}

\newcommand{\Kh}{K\"{a}hler}
\newcommand{\beq}{\begin{equation}}
\newcommand{\eeq}{\end{equation}}
\newcommand{\barr}{\begin{array}}
\newcommand{\earr}{\end{array}}
\newcommand{\equ}[1]{\begin{gather} #1 \end{gather}}

\newcommand{\tabu}[2]{\begin{tabular}{#1} #2 \end{tabular}}

\newcommand{\pmtrx}[1]{\begin{pmatrix} #1 \end{pmatrix}}

\newcommand{\non}{\nonumber}
\newcounter{oldcounter}

\newcommand{\bder}{\bar\partial}
\newcommand{\fA}{\mathfrak{ A}}

\newcommand{\fF}{\mathfrak{ F}}

\newcommand{\fR}{\mathfrak{ R}}

\newcommand{\bee}{{\bar e}}

\newcommand{\bx}{{\bar x}}
\newcommand{\byy}{{\bar y}}
\newcommand{\bz}{{\bar z}}

\newcommand{\bE}{{\bar E}}

\newcommand{\bP}{{\bar P}}

\newcommand{\bge}{{\bar\epsilon}}

\newcommand{\tgch}{{\tilde \chi}}

\newcommand{\Intr}{\mathbb{Z}}
\newcommand{\Cplx}{\mathbb{C}}

\newcommand{\CP}{\mathbb{CP}}

\newcommand{\ba}[2]{\[\begin{array}{#2}\label{#1}}
\newcommand{\ea}{\end{array}\]}
\newcommand{\be}{\begin{equation}}
\newcommand{\ee}{\end{equation}}
\newcommand{\bea}{\begin{eqnarray}}
\newcommand{\eea}{\end{eqnarray}}

\newcommand{\E}[1]{\mathrm{E_{#1}}}
\newcommand{\U}[1]{\mathrm{U(#1)}}
\newcommand{\SU}[1]{\mathrm{SU(#1)}}
\newcommand{\SO}[1]{\mathrm{SO(#1)}}

\newcommand{\rep}[1]{\mathbf{#1}}
\newcommand{\crep}[1]{\overline{\rep{#1}}}

\newcommand{\sm}{{\,\mbox{-}}}

\def\tbZtworesMod{
\begin{table}
\begin{center} 
\begin{tabular}{| c | c | l |}
\hline & &\\ [-2ex]
Model & $G_{\rm blowup}$  & Representations of Hypers
\\[1ex]\hline\hline && \\ [-2ex]
St.\ Emb. & $\SO{28} \times \SU{2}$ &
$\frac {5}8 (\rep{28},\rep{2})+ \frac{45}{16} (\rep{1},\rep{1})$ 
\\[1ex]\hline & & \\ [-2ex]
$(0^{13},1^2,2)$ & $\SO{26} \times \U{2} \times \U{1}$ & 
$ \frac 18 (\rep{26}, \rep{2})_1 + \frac 18 (\rep{1}, \rep{2})_1
+ \frac 78 (\rep{1}, \rep{1})_2 + \frac {7}8 (\rep{26}, \rep{1})_2 
+ \frac {17}8 (\rep{1}, \rep{2})_3 $
\\[1ex]\hline&&\\ [-2ex]
$(0^{10}, 1^6)$ & $\SO{20} \times \U{6}$ &
$ \frac 18 (\rep{20}, \rep{6})_1 + \frac 78 (\rep{1}, \rep{15})_2 $
\\[1ex]\hline & & \\ [-2ex]
$({\frac 12}^{15}, \sm\frac 32)$ & $\U{15} \times \U{1}$ & 
$ \frac 18 (\rep{15})_1 + \frac 18 (\rep{105})_1 + \frac 78 (\rep{15})_2 $
\\[1ex]\hline
\end{tabular} 
\end{center}
\captn{\label{tb:Z2resMod}
The spectra of the standard embedding and the three $\U{1}$ gauge
bundle models on the resolution of the orbifold $\Cplx^2/\Intr_2$ with
vanishing $\cH$ are displayed. The $\U{1}$ charges are the eigenvalues
of $H_V\,$, and determine the multiplicities via~\eqref{multiZ2}. 
}
\end{table}
}

\def\tbZthreeresMod{
\begin{table}
\begin{center} 
\begin{tabular}{| c | c | l |}
\hline && \\ [-2ex]
Model & $G_{\rm blowup}$ & Representations    
\\[1ex]\hline\hline && \\ [-2ex]
St.\ Emb. & $\SO{26} \times \U{1}$ &
$\frac {12}9 (\rep{26})_1 + \frac{12}9 (\rep{1})_{-2}$   
\\[1ex]\hline && \\ [-2ex]
$(0^{12},1^3,3)$ & $\SO{24} \times \U{3} \times \U{1}$ &
$\frac 19 (\rep{24},\rep{3})_{1} + \frac 29 (\rep{1},\rep{3})_{\sm 2}  
+ (\rep{24},\rep{1})_{\sm 3}
+ \frac {26}9 (\rep{1}, \crep{3})_{\sm 4}$
\\[1ex]\hline&& \\ [-2ex]
$(0^{13}, 2^3)$ & $\SO{26} \times \U{3}$ & 
$ \frac 19 (\rep{26},\crep{3})_{\sm 2} + \frac {26}9
(\rep{1},\rep{3})_{\sm 4}$  
\\[1ex]\hline & &\\ [-2ex]
$(0^{10},1^4,2^2)$ & $\SO{20} \times \U{4} \times \U{2}$ &
$ \frac 19 (\rep{1},\crep{4},\rep{2})_1 
+ \frac 19 (\rep{20},\rep{4},\rep{1})_1 
+ \frac 19 (\rep{1},\crep{6},\rep{1})_{\sm 2} $
\\[1ex] &&\\ [-2ex]
&& 
$+ \frac 19 (\rep{20},\rep{1},\crep{2})_{\sm 2} 
+ (\rep{1},\crep{4},\crep{2})_{\sm 3} 
+ \frac {26}9 (\rep{1},\rep{1},\rep{1})_{\sm 4}  $
\\[1ex]\hline & &\\ [-2ex]
$(0^{7}, 1^8,2)$ & $\SO{14} \times \U{8} \times \U{1}$ &
$ \frac 19 (\rep{1}, \crep{8})_1 + \frac 19 (\rep{14}, \rep{8})_1 
+ \frac 19 (\rep{1},\crep{28})_{\sm 2}
+ \frac 19 (\rep{14}, \rep{1})_{\sm 2} 
+ (\rep{1}, \crep{8})_{\sm 3} $
\\[1ex]\hline & & \\ [-2ex]
$(0^{4}, 1^{12})$ & $\SO{8} \times \U{12}$ &
$ \frac 19 (\rep{8}, \rep{12})_1 
+ \frac 19 (\rep{1}, \crep{66})_{\sm 2} $
\\[1ex]\hline & & \\ [-2ex]
$({\frac 12}^{14}, \frac 32,\sm\frac 52)$ & $U(14)\times U(1)\times U(1)$ &
$ \frac 19 (\crep{14})_1 + \frac 19 (\rep{1})_1 + \frac 19 (\rep{91})_1 
+ \frac 19 (\rep{14})_{\sm 2} $ 
\\[1ex] & & \\ [-2ex]
&& 
$ + \frac 19 (\crep{14})_{\sm 2} 
+ (\crep{14})_{\sm 3} + \frac {26}9 (\rep{1})_{\sm 4} $
\\[1ex]\hline & & \\ [-2ex]
$({\frac 12}^{12}, {\frac 32}^4)$ & $\U{4} \times \U{12}$ &
$ \frac 19 (\rep{4},\crep{12})_1 + \frac 19 (\rep{1}, \rep{66})_{1}
+ \frac 19 (\crep{4}, \crep{12})_{\sm 2} + ( \crep{6},\rep{1})_{\sm 3}$
\\[1ex]\hline
\end{tabular} 
\end{center}
\captn{\label{tb:Z3resMod}
This table gives the spectra of the standard embedding and the seven
$\U{1}$ gauge bundle models on the resolution of the orbifold
$\Cplx^3/\Intr_3$ with vanishing $\cH$. The charges are the
eigenvalues of the operator $H_V$, and determine the multiplicities
according to~\eqref{NVZ3}. 
}
\end{table}
}

\def\tbZthreeresE{
\begin{table}
\begin{center} 
\begin{tabular}{| c | c | c |}
\hline && \\ [-2ex]
Model & $G_{\rm blowup}$ & $G_{\rm blow\ down}$
\\[1ex]\hline\hline && \\ [-2ex]
St.\ Emb. & 
$\E{8} \times \E{6}'$ &
$\E{8} \times \E{6}' \!\times\! \SU{3}'$ 
\\[1ex]
&& \\ [-2ex]
\tabu{c}{
$(0^{8}; 0^4,1^3,3)$, $(0^{8};0^5, 2^3)$
\\[1ex] 
 $(0^{8};0^2,1^4,2^2)$
}
& 
$\E{8} \times \SO{10}' \!\times\! \U{3}'$ & 
\\[1ex]\hline && \\ [-2ex]
\tabu{c}{
$(0^{2},1^6;0^5, 1^2,2)$,
$(\frac 12^{6}, \frac 32^2; \frac 12^{6}, \frac 32^2)$ \\[1ex]  
$(0^{5},1^2,2; 0^5,1^2,2)$
}
& 
$\E{6} \!\times\! \U{2} \times \E{6}' \!\times\! \U{2}'$ &
$\E{6} \!\times\! \SU{3} \times \E{6}' \!\times\! \SU{3}'$ 
\\[3ex]\hline & &\\ [-2ex]
\tabu{c}{
$(0^{6},1^2; 0^6,1,3)$, 
$(\frac 12^{8}; \frac 12^{4}, \frac 32^4)$ \\[1ex]
$(0^{6},1^2;0, 1^6,2)$, 
$(\frac 12^{8}; \frac 12^{6}, \frac 32,\sm\frac 52)$ \\[1ex]
$(0^{6},1^2; 0^4,1^2,2^2)$ 
}
& 
$\E{7} \!\times\! \U{1} \times \SO{12}' \!\times\! \U{1}'{}^2$ & 
$\E{7} \!\times\! \U{1} \times \SO{14}' \!\times\! \U{1}'$ 
\\[5ex]
&& \\ [-2ex]
\tabu{c}{
$(0^{7},2;0^6, 2^2)$, 
$(1^{8}; 0^7,2)$ 
}
& 
$\E{7} \!\times\! \U{1} \times \SO{14}' \!\times\! \U{1}'$ & 
\\[1ex]
& &\\ [-2ex]
\tabu{c}{ $(0^6,2^2; 0^{4},1^4)$ } & 
$ \SO{12} \!\times\! \U{2} \times\SO{14}' \!\times\! \U{1}'$ & 
\\[1ex]\hline 
& &\\ [-2ex]
\tabu{c}{ 
$(0^3,1^4,2; 0^{7},2)$, 
$(\frac 12^{7},\frac 52; \frac 12^{7},\sm\frac 32)$ \\[1ex] 
$(0^3, 1^4,2; 0^{4},1^4)$,  
$(\frac 12^{5}, \frac 32^2,\sm\frac 32; \frac 12^{7}, \sm\frac 32)$ 
}
& 
$\U{8} \times \SO{14}' \!\times\! \U{1}'$ & 
$\SU{9} \times\SO{14}' \!\times\! \U{1}'$
\\[3ex]\hline
\end{tabular} 
\end{center}
\captn{\label{tb:Z3resE8}
This table lists all the possible $\E{8}\times \E{8}'$ shifts that
satisfy the integrated Bianchi identity~\eqref{IntBianchiZ3}. We give 
the gauge groups on the resolution of  $\Cplx^3/\Intr_3$ and in the
blow down limit.}
\end{table}
}

\def\tbsuumodels{
\begin{table}
\begin{center} 
\begin{tabular}{| c | c | l | l |}
\hline &&& \\ [-2ex]
$k$ & Shift Vector &$G_{\rm blowup}$ & $G_{\rm blow\ down}$
\\[1ex]\hline &&& \\ [-2ex]
$2$ & $(0^{12},1^2,2,3)$ & 
$\SO{20} \times \U{2} \times \U{1} \times \U{1}$ & 
$\SO{26} \times \U{3}$ 
\\[1ex]
&&& \\ [-2ex]
& $(0^9,1^6,3)$ & 
$\SO{14} \times \U{6} \times \U{1}$ & 
$\SO{20} \times \U{6}$ 
\\[1ex]
&&& \\ [-2ex]
& $(0^{10},1^3,2^3)$ & 
$\SO{16} \times \U{3} \times \U{3}$ & 
\\[1ex]
&&&\\ [-2ex]
& $(0^7,1^7,2^2)$ & 
$\SO{10} \times \U{7} \times \U{2}$ & 
$\SO{14} \times \U{9}$ 
\\[1ex]
&&&\\ [-2ex]
& $(0^4,1^{11},2)$ & 
$\SO{4} \times \U{11} \times \U{1}$ & 
$\SO{8} \times \U{12}$ 
\\[1ex]\hline &&&\\ [-2ex]
$4$ & $(0^{14},3^2)$ & 
$\SO{20} \times \U{2}$ & 
$\SO{32}$ 
\\[1ex]
&&& \\ [-2ex]
& $(0^{13},1^2,4)$ & 
$\SO{18} \times \U{2} \times \U{1}$ & 
$\SO{26} \times \U{3}$ 
\\[1ex]
&&& \\ [-2ex]
& $(0^{12},1,2^2,3)$ & 
$\SO{16} \times \U{1} \times \U{2} \times \U{1}$ & 
\\[1ex]
&&&\\[-2ex]
& $(0^9,1^5,2,3)$ & 
$\SO{10} \times \U{5} \times \U{1} \times \U{1}$ & 
$\SO{20} \times \U{6}$  
\\[1ex]
&&& \\[-2ex]
& $(0^{10},1^2,2^4)$ & 
$\SO{12} \times \U{2} \times \U{4}$ & 
\\[1ex]\hline &&& \\ [-2ex]
$6$ & $(0^{13},1,2,4)$ & 
$\SO{14} \times \U{1}  \times \U{1}  \times \U{1}$ & 
$\SO{26} \times \U{3}$ 
\\[1ex]\hline
\end{tabular} 
\end{center}
\captn{\label{tb:su2u1models}
The first column gives the number of times we have used the $\SU{2}$
bundle~\eqref{SUn-1bundle}. The second column gives the vectorial
shift vector $V$ of these $\SU{2}$--$\U{1}$ bundle models. The final
two columns give the resulting gauge groups on the resolution and in
the blow down limit.  
}
\end{table}
}

\def\tbZthreeorMod{
\begin{table}
\begin{center} 
\begin{tabular}{| c | c | c | c | c |}
\hline &&&& \\ [-2ex]
Orbifold  & Blowup & $G_{\rm orbifold}=$ & Matter spectrum on the  & {Additional}  \\ [0ex]
shift & shift & $G_{\rm blow\ down}$ 
& orbifold resolution & twisted matter 
\\[1ex]\hline &&&&\\ [-2ex]
$(0^{13},1^2,2)$  & $(0^{12},1^3,3)$ & 
$\SO{26} \times \U{3}$ & 
$\frac 19 (\rep{26},\rep{3}) + \frac{26}{9} (\rep{1},\crep{3})  
+ (\rep{26},\rep{1})$& $(\rep{1},\rep{1})$
\\[1ex] &&&&\\ [-2ex]
& $(0^{13},2^3)$  & 
&
$\frac 19 (\rep{26},\crep{3}) + \frac{26}{9} (\rep{1},\rep{3})$
& $(\rep{1},\rep{1})+(\rep{26},\rep{1})$
\\[1ex]\hline &&&&\\ [-2ex]
 $(0^{10},1^4,2^2)$ & $(0^{10},1^4,2^2)$ & 
$\SO{20} \times \U{6}$ &
$ \frac {10}9 (\rep{1},\crep{15}) 
+ \frac 19 (\rep{20},\rep{6}) 
+ 3 (\rep{1}, \rep{1})$& 
\\[1ex]\hline &&&& \\ [-2ex]
$(0^{7},1^6,2^3)$ & $(0^{7},1^8,2)$ & 
$\SO{14} \times \U{9}$ &
$  \frac 19 (\rep{14}, \rep{9})
+ \frac 19 (\rep{1},\crep{36})
+ (\rep{1}, \crep{9}) $ & 
\\[1ex]\hline &&&& \\ [-2ex]
$(0^{4},1^8,2^4)$ & $(0^{4},1^{12})$ & 
$\SO{8} \times \U{12}$ &
$ \frac 19 (\rep{8}, \rep{12})
+ \frac 19 (\rep{1}, \crep{66})$&
$(\rep{1},\rep{1})+(\rep{8_+},\rep{1})$
\\[1ex] &&&& \\ [-2ex]
& $(\frac 12^{12},\frac 32^{4})$ & 
&
$ \frac 19 (\rep{8}, \crep{12})
+ \frac 19 (\rep{1}, \rep{66})+(\rep{8_+},\rep{1})$
& $(\rep{1},\rep{1})$
\\[1ex]\hline &&&& \\ [-2ex]
$(0^{1},1^{10},2^5)$ & $\!(\frac 12^{14},\frac 32, \sm\frac 52)$ & 
$\SO{2}\times \U{15}$ &
$ \frac {11}9 (\rep{15})  
+ \frac 19 (\crep{105})+ 3(\rep{1}) $ 
&
\\[1ex]\hline 
\end{tabular} 
\end{center}
\captn{\label{tb:Z3orMod}
The first column displays the different heterotic $\Intr_3$ orbifold
shifts. The shifts characterizing the $\U{1}$ bundle  models on the
blowup $\cM$ are given in the second column. The gauge groups of the
heterotic orbifold models coincide with the gauge groups of the resolution
models in blow down; they are listed in the next column. The one but
last column gives the matter representations on the resolution. The
last column gives the  {\em additional} twisted matter that the
heterotic string predicts for these orbifold models. 
}
\end{table}
}

\def\fgregdelta{
\begin{figure}
\begin{center} 
\begin{tabular}{ c c c }
\includegraphics[width=7.0cm]{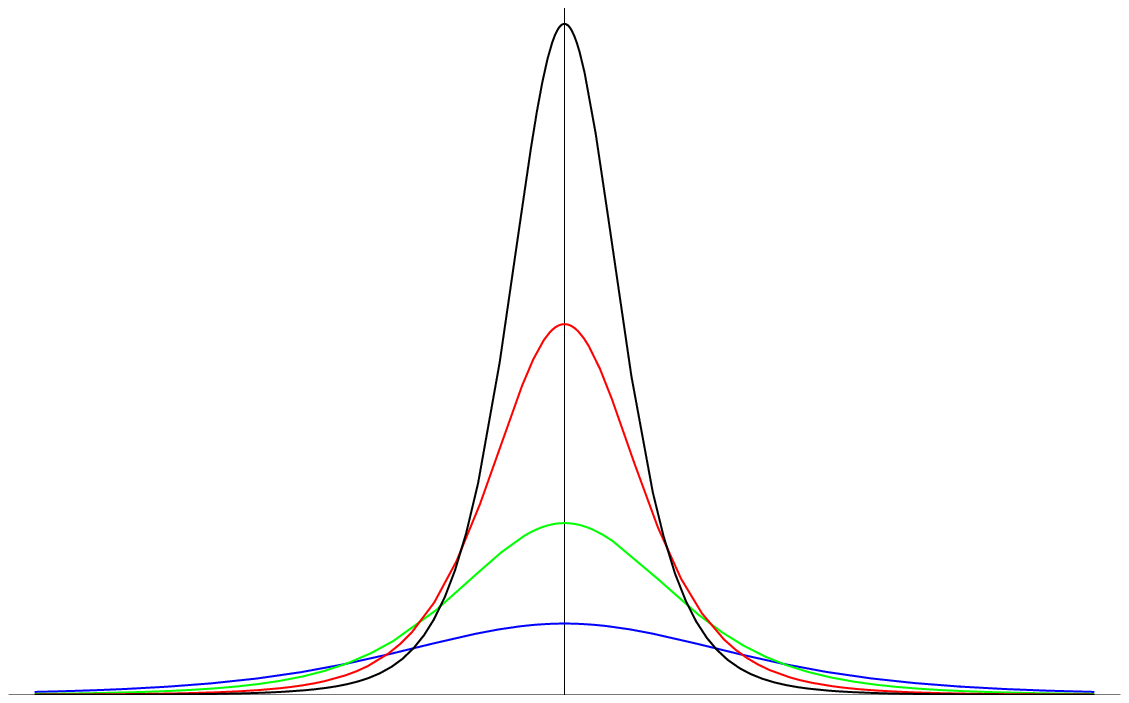} 
& \qquad & 
\includegraphics[width=8.0cm]{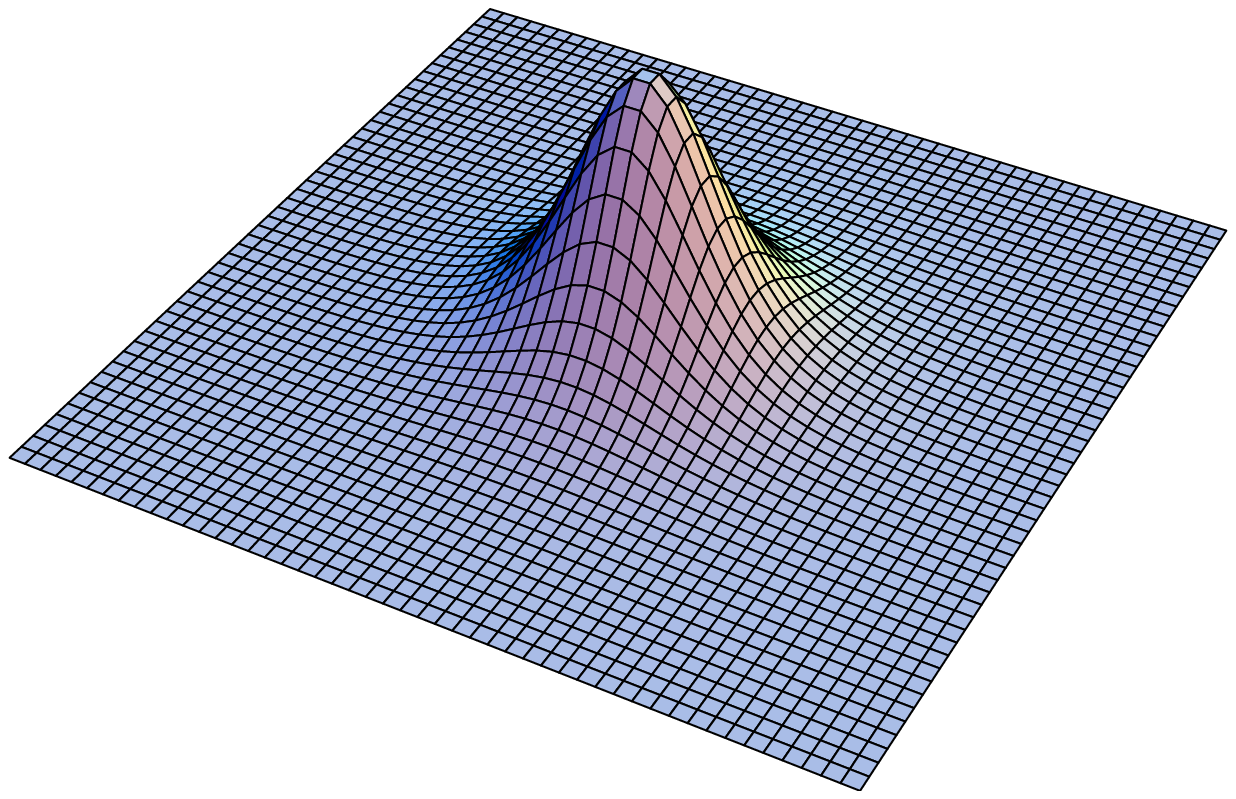} 
\end{tabular} 
\end{center}
\captn{\label{fg:regdelta}
The first picture gives the cross section profiles of the regularized
delta function defined in~\eqref{regdelta} for various values of
$r\,$. The second picture gives a three dimensional impression of its
shape.  
}
\end{figure}
}

\begin{document}

\thispagestyle{empty}

\begin{flushright}
HD-THEP-07-03 \\
SIAS-CMTP-07-1 \\ 
hep-th/0701227\\
\end{flushright}
\vskip 2 cm
\begin{center}
{\Large {\bf Resolutions of $\boldsymbol{\Cplx^n/\Intr_n}$ Orbifolds, their $\boldsymbol{U(1)}$ Bundles, 
\\[1ex]
and Applications to String Model Building} 
}
\\[0pt]

\bigskip
\bigskip {\large
{\bf S.\ Groot Nibbelink$^{a,b,}$\footnote{
{{ {\ {\ {\ E-mail: grootnib@thphys.uni-heidelberg.de}}}}}}},
{\bf M.\ Trapletti$^{a,}$\footnote{
{{ {\ {\ {\ E-mail: M.Trapletti@thphys.uni-heidelberg.de}}}}}}},
{\bf M.G.A.\ Walter$^{c,d,}$\footnote{
{{ {\ {\ {\ E-mail: martin.walter@simon-kucher.com}}}}}}}
\bigskip }\\[0pt]
\vspace{0.23cm}
${}^a$ {\it 
Institut f\"ur Theoretische Physik, Universit\"at Heidelberg, \\ 
Philosophenweg 16 und 19,  D-69120 Heidelberg, Germany 
\\} 
\vspace{0.23cm}
${}^b$ {\it 
Shanghai Institute for Advanced Study, 
University of Science and Technology of China,\\ 
99 Xiupu Rd, Pudong, Shanghai 201315, P.R.\ China
 \\} 
\vspace{0.23cm}
${}^c$ {\it 
Physikalisches Institut der Universit\"at Bonn, \\
Nussallee 12, 53115 Bonn, Germany
 \\} 
\vspace{0.23cm}
${}^d$ {\it 
Simon - Kucher \& Partners, Strategy and Marketing Consultants,  \\ 
Haydnstr. 36, 53115 Bonn, Germany
 \\} 
\bigskip
\end{center}

\subsection*{\centering Abstract}

We describe blowups of $\Cplx^n/\Intr_n$ orbifolds as complex line
bundles over $\CP^{n-1}\,$. We construct some gauge bundles
on these resolutions. Apart from the standard embedding, we describe
$\U{1}$ bundles and an $\SU{n\!-\!1}$ bundle. Both blowups and their
gauge bundles are given explicitly. We investigate ten dimensional
$\SO{32}$ super Yang--Mills theory coupled to supergravity on these
backgrounds. The integrated Bianchi identity implies that there are
only a finite number of $\U{1}$ bundle models. We describe how the
orbifold gauge shift vector can be read off from the gauge
background. In this way we can assert that in the blow down limit 
these models correspond to heterotic $\Cplx^2/\Intr_2$ and
$\Cplx^3/\Intr_3$ orbifold models. (Only the $\Intr_3$ model with
unbroken gauge group $\SO{32}$ cannot be reconstructed in blowup
without torsion.) This is confirmed by computing the charged chiral
spectra on the resolutions. The construction of these blowup models
implies that the mismatch between type--I and heterotic models on
$T^6/\Intr_3$ does not signal a complication of $S$--duality, but
rather a problem of type--I model building itself: The standard
type--I orbifold model building only allows for a single model on this
orbifold, while the blowup models give five different models in blow down.

\newpage 
\setcounter{page}{1}

\section{Introduction}
\labl{sc:intro}

After it was realized that heterotic string compactifications can
result in chiral models in four dimensions~\cite{Candelas:1985en,Witten:1985bz},
such compactifications have been studied by many authors. These
compactifications require a detailed understanding of Calabi--Yau 
manifolds, but as they are complicated spaces their behavior is still an
active field of study. There has been a  strong effort to obtain
MSSM--like models from the $\E{8}\times\E{8}$ heterotic string by 
compactifying on elliptically fibered Calabi--Yaus with $\SU{n}$ gauge
bundles~\cite{Braun:2005ux,Braun:2005bw}. More general applications of
$\U{1}$ and $\SU{n}$ bundles are discussed
in~\cite{Honecker:2006dt,Honecker:2006qz}  
and~\cite{Andreas:2004ja,Blumenhagen:2005ga,Blumenhagen:2005pm,Weigand:2005ng} for example.

For model building purposes orbifold 
compactifications~\cite{dixon_85,Dixon:1986jc,Ibanez:1988pj}
proved very useful, because they capture all the stringy features, while
at the same time are completely calculable. The number of possible
$T^4/\Intr_n$ and $T^6/\Intr_n$ models with multiple Wilson lines is
very large. (For lengthy lists of orbifold models see
e.g.~\cite{Casas:1989wu,Katsuki:1989kd,Katsuki:1990bf,Kobayashi:1991mi,Kobayashi:1994rp,Kawamura:1996zu}.)
Most works on orbifold compactifications have focused on the heterotic
$\E{8}\times\E{8}$ string, surprisingly late also orbifolds of
the $\SO{32}$ heterotic string have been
considered~\cite{Giedt:2003an,Choi:2004wn,Nilles:2006np}. 
Even though the number of models with various Wilson lines is very
large, their properties at the various fixed points can be easily
understood. At a given fixed point the spectrum and properties are
the same as those at a fixed point of a pure orbifold model with an
appropriately chosen gauge shift vector. These so--called fixed point
equivalent models proved very useful in the analysis of local anomaly
cancellation and $D$--term tadpoles in heterotic 
orbifolds~\cite{Gmeiner:2002es,GrootNibbelink:2003gb,Nibbelink:2003rc}.

Orbifolds were initially considered as simple prototypes
of Calabi--Yau compactifications, the exact relation between them is
mostly understood on the topological level: The orbifold singularities
can be cut out and replaced by Eguchi--Hanson spaces. In this
way some topological properties of singularities can be understood. 
Also the study of anomalies and tadpoles at singularities has shown
that many properties are determined by the local geometry
only. Therefore, to understand the behavior of blowups of orbifolds it
can often be sufficient to perform a resolution analysis at a single
fixed point only. Using toric geometry substantial progress has been
made to understand the topological properties of blowups of orbifolds
in a systematic way, see e.g.~\cite{Lust:2006zh}.

In this work we would like to go beyond a purely topological
description, and obtain the geometrical objects like metric and
curvature of the Calabi--Yau resolution of orbifold singularities
explicitly. For simplicity we consider the orbifolds of the type 
$\Cplx^n/\Intr_n\,$, $n\geq 2\,$, only. The orbifold $\Cplx^2/\Intr_2$
is also known as the conifold. These Eguchi--Hanson
spaces~\cite{Eguchi:1978xp} are well--known,
see~\cite{pol_2,Joyce:2000,Cvetic:2001zb} for example. The procedure,
we use to obtain these non--compact Calabi--Yaus, is similar to the
method explained in~\cite{Candelas:1989js} to derive the metric of the
resolved conifold (see also e.g.~\cite{PandoZayas:2000sq}). 
Non--compact Calabi--Yaus in six real dimensions with a $\CP^2$ base
were obtained in~\cite{Page:1985bq,Berard-Bergery:1986}. The \Kh\
potentials for resolutions of $\Cplx^n/\Intr_n$ are given
in~\cite{Calabi:1979}. Our construction uses some properties of \Kh\
coset spaces and is closely related
to~\cite{Higashijima:2002px,Higashijima:2001fp,Higashijima:2001vk}.  
(For resolutions of codimension two singularities see for
example~\cite{Serone:2004yn,Wulzer:2005ps}.) Moreover, we would like
to explicitly construct gauge backgrounds on these resolutions, that
satisfy the Hermitian Yang--Mills equations. Once we have both the
geometrical and gauge backgrounds in hand, we can simply compute
various integrals, that are relevant for consistency requirements
and that determine the spectra of models at orbifold resolutions. We
use anomaly cancellation and comparison with the spectra of heterotic
$T^4/\Intr_2$ and $T^6/\Intr_3$ orbifold models as checks of
the validity of this procedure.

The paper is structured as follows: Section~\ref{sc:geometry} first
describes the geometry of $\Cplx^n/\Intr_n$ orbifolds using 
coordinates that are useful in the construction of the Ricci--flat
\Kh\ blowup as a complex line bundle over $\CP^{n-1}\,$. This
construction is described in detail relying on some properties of
\Kh\ geometry, and results in explicit formulae for the
spin--connection and the curvature. In section~\ref{sc:u1bundles} we
first explain how orbifold boundary conditions of gauge fields can be
reformulated as contour integrals around the blowup of the singularity
in the blow down limit.  We then give a number of examples of gauge
bundles, that can be matched with orbifold boundary conditions in this
way. These examples consist of the standard embedding, $\U{1}$ and
$\SU{n\!-\!1}$ gauge bundles. Section~\ref{sc:consistent} explains how
the $\U{1}$ bundles can be used to obtain consistent reductions of ten
dimensional super Yang--Mills theory coupled to six and four
dimensions. In particular, we determine all possible gauge shift
vectors of the consistent $\U{1}$ bundles for these resolutions. In
addition, we compute the charged chiral spectra of these models. In
section~\ref{sc:heterotic} we give a detailed account how the spectra
of the blowup models are related to heterotic $\SO{32}$ orbifold
models. Section~\ref{sc:concl} is devoted to the conclusions, explains
some consequences for type--I model building, and 
discusses possible extensions of this work. Appendix~\ref{sc:CPn}
gives some technical details of forms on $\CP^{n-1}$ and its complex
line bundle. In appendix~\ref{sc:integrals} we list a number of
integrals of forms of $\CP^{n-1}$ and the resolution of
$\Cplx^n/\Intr_n\,$, which are frequently used in the main part of the
text.

\section{The Geometry of the Resolution of the $\boldsymbol{\Cplx^n/\Intr_n}$ Orbifold}
\labl{sc:geometry}

In this section we describe explicitly the resolution of the
$\Cplx^n/\Intr_n$ orbifold for arbitrary $n \geq 2\,$. This orbifold
is defined as the complex space $\Cplx^n$ with coordinates $\tZ^A\,$,
on which the $\Intr_n$ twist acts as
\equ{
\gTh (\tZ) ~=~ \gth \, \tZ~, 
\qquad 
\gth ~=~\text{diag} 
\Big( e^{2\pi i\, \gf_1/n}, \ldots, e^{2\pi i\, \gf_n/n}  \Big)~, 
\qquad 
\gf ~=~ \big(1^{n-1}, 1-n \big)~. 
\labl{geomshift}
}
We have chosen the geometrical shift $\gf$ in~\eqref{geomshift} such
that the sum of its entries vanishes. This guarantees that the action
is also well--defined on spinors. Moreover, as this choice allows for
some invariant spinors, it ensures that some supersymmetry is
preserved. This orbifold has an $\SU{n}$ isometry group, that acts by
matrix multiplication as $\tZ \ra g\, \tZ$ for $g \in \SU{n}\,$,
because the  orbifold twist is proportional to the identity on the
bosonic coordinates $\tZ\,$.

One can define $n$ coordinate patches for the resulting orbifold
$\Cplx^n/\Intr_n - \lbrace 0 \rbrace\,$, see
e.g.~\cite{Klerk:Master2002}. In each of them one of the $n$
coordinates is non--vanishing and has a deficit angle $2\gp/n\,$. 
The $n$ coordinate patches are all equivalent and related to each
other by $\SU{n}$ transformations. Since the orbifold is flat (apart
from the singular point) complex manifold, it can be described by the 
standard \Kh\ potential of $\Cplx^n\,$:   
\equ{
\cK_{\Cplx^n/\Intr_n} ~=~ \cK_{\Cplx^n} ~=~ \bar{\tZ} \tZ~.  
}
We now would like to use coordinates that allow for a systematic
construction of resolutions of the orbifold as line bundles over 
$\CP^{n-1}\,$, which are defined as follows: 
Let $z^i$ with $i = 1,\dots, n\!-\!1$ 
be local coordinates of $\CP^{n-1}=\SU{n}/\U{n-1}$ then 
we may write\footnote{More precisely, $z$ is a local coordinate 
of the complexified coset $\SU{n}^{\Cplx}/\hat{\mathrm{U}}(n-1)\,$, 
as has been extensively discussed in~\cite{Bando:1984ab}.}
\equ{
\tZ ~=~ \gx(z) \pmtrx{ 0_{n-1} \\ \tZ^n}~, 
\qquad 
\gx(z) ~=~ \pmtrx{ \Id_{n-1} & z \\ 0 & 1}~,
\labl{ParaOrbi}
}
in the coordinate patch where $\tZ^n \neq 0$ has the $2\gp(1-1/n)$
deficit angle. One can introduce a new complex variable  
$x = (\tZ^n)^n\,$,  which does not have a deficit angle, i.e.\ 
$0 < \arg(x) < 2\gp\,$. The \Kh\ potential becomes 
\equ{
\cK_{\Cplx^n/\Intr_n} ~=~ X^{\frac 1n}~, 
\qquad 
X = \bx \, \gch^n\, x~, 
\qquad 
\gch = 1 + \bz z~: 
\labl{KahlOrbi}
}
The deficit angle has been replaced by a non--analyticity in the
\Kh\ potential. This expression of the \Kh\ potential still manifestly
possesses all $\SU{n}$ isometries, because it is written in terms of
the variable $X\,$, which is invariant under them. We will use this
\Kh\ potential to show that, in an appropriate limit  the resolution
$\cM^n\,$, tends to the orbifold $\Cplx^n/\Intr_n\,$.

We now proceed to define this blowup of $\Cplx^n/\Intr_n$ by
constructing a cone over $\CP^{n-1}\,$. The cone is defined as
the $n$th power of the fundamental complex line bundle over
$\CP^{n-1}\,$. (For a detailed discussion of the \Kh\ geometry of
$\CP^{n-1}$ and its complex line bundles,
see~\cite{GrootNibbelink:2000gq,GrootNibbelink:2000hu,GrootNibbelink:1999un}.)
This cone itself is a \Kh\ manifold but in general it is not
Ricci--flat. By requiring Ricci--flatness we obtain the resolution
manifold $\cM^n\,$, that we want to obtain. Similar constructions of \Kh\
cones on $\CP^{n-1}$ and more general coset spaces can be found 
in~\cite{Higashijima:2002px,Higashijima:2001fp,Higashijima:2001vk}. 
By requiring that the resolution has the  full $\SU{n}$ isometries of the 
orbifold, its geometry is uniquely defined by its  \Kh\ potential 
\equ{
\cK ~=~ \cK(X)~, 
\labl{Kahl}
}
as a function of the variable $X\,$,defined in~\eqref{KahlOrbi}, only. 
The resulting \Kh\ metric 
\equ{
G ~=~ \pmtrx{\dsp 
n \, M\, \gch\inv\, \tgch\inv 
+ M' \gch^n \, (n\, {z \gch\inv \bx}) (n\, {x \gch\inv\bz}) 
&\dsp  
M' \gch^n \, (n\,{z \gch\inv \bx}) 
\\[2ex]\dsp 
M' \gch^n \, (n\, {x \gch\inv \bz})
& 
M'\gch^n
}~, 
\labl{Metric} 
}
with the $n\!-\!1 \times n\!-\!1$ matrix  $\tgch = \Id_{n-1} +z\bz\,$,
depends on the combination $M(X) = X \cK'(X) $   
involving the first derivative $\cK'(X)$ of $\cK(X)$ w.r.t.\ $X$ only.

To obtain the non--compact Calabi--Yau manifold $\cM^n$ we 
enforce the Ricci--flatness condition following~\cite{Candelas:1989js}: 
The Ricci--tensor $R_{\uA A}$ of a \Kh\ manifold is given by   
\equ{
R_{\uA A} ~=~ [\ln \det G]_{,\uA A}~. 
}
Therefore, to obtain a Ricci--flat manifold the determinant $\det G$
has to factorize into purely holomorphic and anti--holomorphic
parts, i.e.\ $\det G = P(z,x)\bP(\bz,\bx)\,$. The determinant of the 
metric of cone \eqref{Metric} takes a surprisingly simple form 
\equ{
\det G ~=~ n^{n-1} \, M^{n-1} M'~. 
\labl{detG}
} 
Since neither $M$ nor $M'$ factorize, the Ricci--flatness implies that
$\det G$ is a constant. Hence we obtain a first order ordinary 
differential equation for $M$, i.e.\ second order ordinary differential
equation for $\cK\,$. The expression for the \Kh\ potential~\eqref{Kahl}
is uniquely determined by two integration constants and the constant
value of the determinant $\det G\,$. Since a \Kh\ potential of a
manifold is only defined upto holomorphic and anti--holomorphic
functions, the last integration constant is irrelevant. An additional
relation between these constants is found by demanding, 
that there is a blow down limit in which the cone tends to the orbifold
$\Cplx^n/\Intr_n\,$, i.e.\ $\cK$ tends to~\eqref{KahlOrbi}. The
remaining variable we call $r$ and the resulting \Kh\ potential is
given by   
\equ{
\cK(X) ~=~ 
\int\limits_1^X \frac {\d X'}{X'} \, M(X')~, 
\qquad 
M(X) ~=~ \frac 1{n} \big( r + X \big)^{\frac 1n}~. 
\labl{FunK}
}
The constant lower bound of the integral is irrelevant as stated
above, because all physics in the end depends on the metric. The blow
down of the resolution is given by the limit $r \ra 0\,$.

Since we have the explicit resolution of the orbifold, it is
interesting to see what happens to the spin--connection and the
curvature  in the blow down limit. To facilitate the discussion of
gauge bundles on this space in the next section, we employ form
notation. In this language the metric can be decomposed into holomorphic
and anti--holomorphic vielbein 1--forms $E$ and $\bE$ as   
\equ{
G ~=~ \bE \otimes E~, 
\qquad 
E ~=~ \pmtrx{ \sqrt{n M} \, e \\[1ex] \sqrt{M'} \, \ge }~. 
}
Here the holomorphic vielbein $e$ of $\CP^{n-1}\,$ is a vector of
$n\!-\!1$ 1--forms, and $\ge$ is a 1--form associated with a complex
line bundle. Their explicit expressions read 
\equ{
e ~=~  \gch^{-\frac 12} \tgch^{-\frac 12}\,  \d z~, 
\qquad 
\ge ~=~ \d y ~+~ n \, i \cB\, y~,
\labl{CPnvielb}
} 
where $y = \gch^{\frac n2} x$ is a convenient complex variable for the
fiber of the line bundle over $\CP^{n-1}\,$. In addition $i \cB$ is a 
$\U{1}$ connection 1--form obtained by taking the trace of the $\U{n\!-\!1}$
connection 1--form $i \tB$ on $\CP^{n-1}\,$: 
\equ{
i \cB ~=~ - \tr( i\tilde \cB) ~=~  \frac 12 ( \bz \, e - \bee \,z)~, 
\qquad 
i \tilde \cB ~=~ 
\tgch^{-\frac 12} \bder (\tgch^{\frac 12}) 
~-~ \der (\tgch^{\frac 12}) \tgch^{-\frac 12}~. 
\labl{CPnUnconn}
} 
More detailed properties of these $\CP^{n-1}$ forms are collected in
appendix~\ref{sc:CPn}.

The spin connection 1--form $\gO$ and the curvature 2--form
$\cR$ of the blowup $\cM^n$ are defined as usual by 
\equ{ 
\d E ~+~ \gO\, E ~=~ 0~, 
\qquad 
\cR ~=~ \d \gO ~+~ \gO^2~. 
}
In these expressions, and throughout this work, we keep the wedge
products implicit in our notation. Using the 1--forms defined above,
the spin--connection reads 
\equ{
\gO ~=~ \pmtrx{ \dsp 
i (\tilde \cB - \cB) ~+~ \frac 1{2n}\, 
\frac{ \byy \, \ge - \bge\, y }{r + X}
& \dsp 
\frac{ \byy \, e }{\sqrt{r + X}} 
\\[3ex] \dsp 
- \frac{ \bee \, y }{\sqrt{r+X}}
& \dsp 
n\, i \cB ~+~  \frac {n\!-\!1}{2n} \, 
\frac{ \byy \, \ge - \bge\, y }{r+X}
}~, 
\labl{SpinConn} 
}
and the curvature 2--form becomes 
\equ{
\cR ~=~ \frac{r}{r + X} \, 
\pmtrx{ \dsp 
 e \, \bee ~-~ \bee \, e  
~+~ \frac 1{n}\, \frac{ \bge \, \ge}{r + X}
& \dsp 
\frac{\bge \, e} {\sqrt{r + X}} 
\\[3ex] \dsp 
\frac{\bee\, \ge}{\sqrt{r + X}} 
& \dsp 
n\, \bee \, e ~-~ \frac {n\!-\!1}{n} \,  \frac{ \bge \, \ge}{r + X}
}~.
\labl{Curv2expl}
}
It is not difficult to check that both the spin--connection and the
curvature are traceless, i.e.\ they are $\SU{n}$ algebra
elements. This means that the manifold $\cM^n$ has $\SU{n}$ holonomy.

As a simple application of the explicit form of the curvature
in~\eqref{Curv2expl}, we compute the Euler numbers of the resolutions 
$\cM^2$ and $\cM^3$ directly. Using that the Euler number
$\gch(\cM^n)$ can be computed by integrating the Euler class
$e(\cM^n)$ (see e.g.~\cite{Nakahara:1990th}), we find 
\equ{
\gch(\cM^2) ~=~ 
\frac 12 \int_{\cM^2} \tr \Big( \frac{\cR}{2\pi i} \Big)^2
~=~ - \frac 32~, 
\qquad 
\gch(\cM^3) ~=~ 
\frac 13 \int_{\cM^3} \tr \Big( \frac{\cR}{2\pi i} \Big)^3
~=~ - \frac 83~, 
\labl{EulerNumbers}
}
see the integrals~\eqref{trR2int} and~\eqref{trR3int} in
appendix~\ref{sc:integrals}. These numbers can be confirmed by the
following consistency checks: $K3$ can be viewed as the blowup of
$T^4/\Intr_2\,$. This orbifold has $16$ fixed points. Each fixed point
can be replaced by the resolution $\cM^2$, hence the Euler number of 
$K3$ is $-24\,$, confirming the well--known result. Similarly, it is
known that the Euler number of the blowup of $T^6/\Intr_3$ is $-72\,$,
see~\cite{pol_2}. This is also consistent with~\eqref{EulerNumbers},
because $T^6/\Intr_3$ has $27$ fixed points.

\fgregdelta

Clearly both spin connection and curvature are regular functions
of the coordinates for any non--zero value of the resolution parameter
$r>0\,$. For any non--zero $x$ the curvature tends to zero in the
blow down limit $r \ra 0\,$. For $x=0$ the space is non--singular; 
the singular point of the orbifold $\Cplx^n/\Intr_n$ is replaced by a
$\CP^{n-1}$ at $x=0\,$. Similarly, for fixed $r>0\,$, the
resolution becomes flat far away from the blown up singularity 
$x\ra\infty\,$. Contrary, if $r=0\,$, we see that parts of the spin
connection and the whole curvature explode in the limit in which $x$
tends to zero. This shows that we can interpret the resolution as a
regularization of the orbifold fixed point delta function. To make
this more precise, consider the two dimensional complex case ($n=2$) for
example, and compute $\tr\, \cR^2\,$. From~\eqref{trR2}
and~\eqref{trR2int} of appendix~\ref{sc:integrals} we conclude that we
can define a regularized orbifold delta function as  
\equ{
\gd_r(z,x) ~=~ \frac1{(n+1)(2\pi)^2} \, \tr\, \cR^2
~=~ -\frac 1{2 \gp^2} \frac {r^2}{(r+X)^3} \bee e\, \bge \ge
~, 
\qquad 
\int_{\cM^2} \gd_r(z,x) ~=~ 1~. 
\labl{regdelta} 
}
In figure~\ref{fg:regdelta} we have made schematic two and three
dimensional pictures of this smeared out delta function in two complex
dimensions.

\section{Gauge Bundles on the Resolution}
\labl{sc:u1bundles}

Next we turn to the construction of non--trivial gauge backgrounds on
the resolution of the orbifold $\Cplx^n/\Intr_n\,$. For simplicity we
consider only the gauge group $\SO{32}\,$. (The extension to the
$\E{8} \times \E{8}'$ gauge group is straightforward, and will be used
at the end of section~\ref{sc:consistent} to classify
$\E{8}\times\E{8}'$ models on the blowup.) We begin with a
short review of gauge theories on orbifolds.

The group $\SO{32}$ is generated by 16 Cartan algebra
elements $H_I\,$, with $I=1,\ldots 16$, and the elements $E_w$
parameterized by the vectorial weights 
$w = (\undr{\pm 1^2, 0^{16}})\,$, with all permutations as the
underline is denoting. These weights are the eigenvalues  
of the commutators 
\equ{
[H_I, E_w] ~=~ w_I \, E_w~. 
}
The gauge field 1--form $i \fA$ takes values in the algebra of
$\SO{32}\,$; by $i\fF$ we denote its field strength. Gauge fields $i
\fA$ on orbifolds can satisfy non--trivial boundary  
conditions 
\equ{
\fA(\gTh\, \tZ) ~=~ U\, \fA(\tZ) \,U\inv~,
\qquad 
U ~=~ e^{2\pi i\, V^I H_I/n}~, 
\labl{OrbiAc} 
} 
for the orbifold action defined in~\eqref{geomshift}. 
In order that this defines a proper $\Intr_n$ action on vectorial
weights, the shift vector $V$ can only contain either integer or only
half--integer entries. (We use a normalization of the gauge shift
vector $V$ without an explicit $\Intr_n$ orbifold factor $1/n\,$.) The
former are called vectorial shifts, and the latter spinorial
shifts. In terms of the coordinates $x = |x|\, e^{i \gvf}$ and $z$, the
orbifold action~\eqref{OrbiAc} takes the form of a periodicity
condition for the angular variable $\gvf$ 
\equ{
\fA(z, |x|, \gvf + 2 \gp) ~=~ 
U\, \fA(z, |x|, \gvf)  \, U\inv~. 
}
By a gauge transformation $g = e^{-i\, \gvf\, V^I H_I/n}$ this
periodicity condition, can be rewritten as 
\equ{
i \fA_g ~=~ g (i \fA + \d ) g\inv ~=~ 
i A + i \cA~, 
\qquad 
i \cA ~=~ i\, \frac 1n\, V^I H_I\,  \d \gvf~, 
}
where $i A$ is a periodic 1--form gauge potential, and $i \cA$ is a 
constant Wilson--line background gauge connection. A gauge invariant 
way of stating that there is a Wilson--line is given by the following
prescription: Consider a loop  $\der C = \{\gvf ~~| ~~0 < \gvf < 2\gp \}$ 
at fixed $z$ and $|x|\,$, and let $C$ represent any surface that has
$\der C$ as it boundary. By Stoke's theorem we have
\equ{
\int_C i \cF
~=~  \int_{\der C} i \cA ~=~ 
2\gp\, i\, 
\frac 1n \, V^I H_I~, 
\labl{IdentifyWilson}
}
where $i\cF$ is the field strength of the $\U{1}$ background $i\cA\,$. This
completes the review of the description of gauge bundles on orbifolds.

We would like to find gauge backgrounds on the resolution of the
orbifold $\Cplx^n/\Intr_n\,$. In order to preserve $N=1$
supersymmetry, the field strength $i\cF$ of the background gauge
potential $i\cA$ has to satisfy the so--called Hermitian Yang--Mills
equations   
\equ{
\cF_{AB} ~=~ 0~, 
\qquad 
\cF_{\uA\, \uB} ~=~ 0, 
\qquad 
\Tr [ \cF] ~\equiv~ G^{A \uA} \, \cF_{\uA A} ~=~ 0~, 
\labl{SusyBack}
}
see \cite{Candelas:1985en}. We study solutions of these equations on 
$\cM^n$ for general $n$ in this section. These solutions should be
regular over the whole manifold $\cM^n$ as long as we have not yet
taken the orbifold limit.

As in the previous section, all forms on $\cM^n$ can be expressed in
terms of the holomorphic 1--forms $e, \ge$ and their
conjugates. Therefore we would like to reformulate these conditions
in terms of these forms: The first two conditions of \eqref{SusyBack}
simply mean that the field strength $i\cF$ only contains mixed
2--forms, like $e \bee\,$, $\bge e$ and $\bge \ge\,$. Taking the last
equation in~\eqref{SusyBack} as the 
definition of the trace of mixed 2--forms, we find   
\equ{
\Tr[ e \bee] ~=~ \frac 1{nM}\, \Id_{n-1}~,
\quad 
\Tr[ \bee e ] ~=~ - \frac {n\!-\! 1}{nM}~, 
\quad 
\Tr[ \ge \bge] ~=~ \frac 1{M'}~, 
\quad 
\Tr[ \bge e] ~=~ \Tr[\bee \ge] ~=~0~,
\labl{Trtrace}
}
in terms of the function $M(X)$ given in \eqref{FunK}. Hence we are
looking for gauge backgrounds which have field strengths that only
contain mixed 2--forms that trace to zero, using the trace defined
by~\eqref{Trtrace}.

In the following we give a few examples of explicit solutions of the
Hermitean Yang--Mills equations on the resolution $\cM^n\,$.  
We determine the corresponding gauge shift vector $V$ on the orbifold
by computing the integral~\eqref{IdentifyWilson} in the blow down
limit. We do not aim to give a complete classification here, but just
construct a number of interesting examples to be considered later.

\subsection{Standard Embedding: An $\boldsymbol{\SU{n}}$ Bundle}

The first example is the well--known standard embedding of the
spin--connection in the gauge bundle: $i \cA_{\rm SE} = \gO$, which is
given in~\eqref{SpinConn}. This is indeed a solution of the Hermitean
Yang--Mills equations as we can see from the field strength $i
\cF_{\rm SE} = \cR\,$: From the expression for $\cR$, 
see eq.~\eqref{Curv2expl}, it follows that it only contains mixed
2--forms, and by a direct computation we find 
\equ{
\Tr[\cR] ~=~0~. 
}
For any arbitrary value of the resolution parameter $r$ this gauge
bundle fills a full $\SU{n} \subset \SO{32}\,$. To determine whether
the standard embedding corresponds to an orbifold Wilson line, 
we compute the integral defined in~\eqref{IdentifyWilson}: 
\equ{
\int_{\der C} \gO ~=~  
\frac {2\gp i}n \, \frac {X}{r + X} \, 
\pmtrx{ \Id_{n-1} & 0 \\ 0 & 1\!-\!n } 
~\ra~ 
\frac{2\gp i}{n} \, 
\pmtrx{ \Id_{n-1} & 0 \\ 0 & 1\!-\!n } 
~. 
\label{geomshiftID}
}
This expression is diagonal, which shows that in the blow down limit
($r \ra 0$)  the standard embedding gives rise to the orbifold
boundary conditions specified by the shift vector 
$V = (1^{n-1},1\!-\!n, 0^{16-n})\,$. Notice that this also 
gives the geometrical shift vector~\eqref{geomshift} back.

\subsection{Construction of $\boldsymbol{\U{1}}$ Background Gauge
  Field}
\labl{sc:U1back}

Next we would like to construct a $\U{1}$ gauge background on the
blowup of the orbifold $\Cplx^n/\Intr_n\,$. A first guess for such a
background is the $\U{1}$ connection $i\cB$ defined
in~\eqref{CPnUnconn}, but this choice does not satisfy the last
condition in~\eqref{SusyBack} required to preserve supersymmetry. In
order to obtain a background that does satisfy this requirement, we
extend the connection as follows
\equ{ 
i \, \cA ~=~ i\, \cB ~+~ e^{-\frac p2} (\bder - \der) e^{\frac p2} 
~=~   i\, \cB + \frac 12 \, p'(X) \, ( \bge y  - \byy \ge)~,
}
where $p(X)$ is an arbitrary function of the $\SU{n}$ isometry
invariant variable $X\,$. As can be seen from the final expression,
only its first derivative $p'(X)$ is of physical relevance. The field
strength 2--form is given by 
\equ{
i\, \cF ~=~   (1- n\, p' X) \bee\, e - (p' X)' \, \bge \, \ge~.  
} 
By computing the trace of this gauge background, we obtain 
\equ{
\Tr[i\cF] ~=~ \frac{\big(p' X M^{n-1} \big)'}{M^{n-1}M'}
~-~ \frac {n-1}{nM}~. 
}
This background is supersymmetric if this trace vanishes, hence we
obtain a simple differential equation for $p'\,$. By solving this
equation, and demanding that the solution is nowhere singular on the
resolution $\cM^n\,$, we determine the $\U{1}$ gauge connection 
\equ{
i \cA ~=~ i\cB ~+~ \frac 1{2n} \frac 1X 
\Big[   
1 ~-~ \Big( \frac r{r +  X }\Big)^{1-\frac 1n}
\Big] \big( \bge y - \byy \ge \big)~, 
\labl{AU1basis}
} 
with field strength 
\equ{
i\cF ~=~ \Big(\frac r{r + X} \Big)^{1-\frac 1n}
\Big( 
\bee e ~-~ \frac {n-1}{n^2} \, \frac 1{r+X}\, \bge \ge
\Big)~, 
\labl{FU1basis}
}
uniquely. Observe that $i\cA$ and $i\cF$ are indeed regular in the
limit $x \ra 0$ for finite values of the resolution parameter
$r\,$. At $x=0$ the field strength diverges in the limit
$r\ra 0\,$. Hence, like the curvature~\eqref{Curv2expl}, it can be
used to define a regularized orbifold fixed point delta function,
similar to the one depicted in figure~\ref{fg:regdelta}.

Using this background, we can easily construct a large class of
$\U{1}$ bundles. In $\SO{32}$ we can embed at most 16 mutually 
commuting $\U{1}$s, precisely parameterizing a Cartan subgroup. 
Using the generators $H_I$ of this Cartan subgroup, we define 
 \equ{
i  \cA_{V} ~=~ i \cA\, V^I H_I~,
\qquad
i \cF_V ~=~ i\, \cF\, V^I H_I~,  
\labl{U1bundle}
} 
where $i\cA$ and $i\cF$ are given in~\eqref{AU1basis}
and~\eqref{FU1basis}, respectively. This bundle is well--defined only
if the first Chern class is integral on all closed 2--cycles for all
relevant representations. By a direct computation we find for the
integral over the $\CP^1$ at $x=0$ 
\equ{
\frac 1{~ 2\gp i}
\int_{\CP^1} i \cF_V ~=~ V^I H_I~, 
\label{Integral1stCh}
}
using~\eqref{CPnInts} of appendix~\ref{sc:integrals}. Therefore, as in
the orbifold case (see below~\eqref{OrbiAc}) the entries of $V$ are
either all integer or all half integer. The same condition is obtained, 
because the gauge background corresponds to orbifold boundary
conditions in the blow down limit: By computing the
integral~\eqref{IdentifyWilson} we find  
\equ{
\int_{\der C} i\cA_V ~=~ 
- \frac {2\gp i}{n} \, V^I H_I\, 
\Big[   
1 ~-~ \Big( \frac r{r +X} \Big)^{1-\frac 1n}
\Big]
~\ra~ 
- \frac {2\gp i}{n}\, V^I H_I~, 
\labl{blowdownU1}
}
in the blow down limit $r \ra 0\,$. This means that the $\U{1}$
bundles on the non--compact Calabi--Yau $\cM^n$ are quantized in units
of $1/n\,$.

\subsection{An $\boldsymbol{\SU{n\!-\!1}}$ Bundle}
\labl{sc:SU2bundle}

The final bundle we describe has an $\SU{n\!-\! 1}$
structure. In~\eqref{CPnUnconn} we obtained a $\U{n\!-\! 1}$ and a
$\U{1}$ bundle on $\CP^{n-1}\,$. By combining them we can obtain an
$\SU{n\!-\! 1}$ gauge connection and field strength  
\equ{
i \tilde \cA ~=~ i \tilde \cB ~+~ \frac 1{n-1}\, i \cB~, 
\qquad
i \tilde \cF ~=~ \d (i \tilde \cA) ~+~ (i \tilde \cA)^2 
~=~ e\, \bee ~+~ \frac 1{n-1}\, \bee\, e~.
\labl{SUn-1bundle}
}
It is not difficult to check that $\tilde \cA$ is indeed an $\SU{n\!-\! 1}$ 
gauge potential, i.e.\ $\tr\, i\tilde \cA = \tr\, i\tilde \cF = 0\,$ (trace over the
external $\SU{n\!-\! 1}$ indices). The field strength is nowhere
vanishing. In addition using the trace of mixed 2--forms defined 
in~\eqref{Trtrace}, we infer that this defines a supersymmetric
background on $\cM^n\,$, because $\Tr[i\tilde \cF] =0\,$. However, 
the integral over $\der C$ is zero, because it
does not contain any $\ge$ or $\bge$ forms. Hence, this configuration
does not correspond to a Wilson line configuration in the orbifold
limit, and cannot be described by a gauge shift vector on the orbifold
$\Cplx^n/\Intr_n\,$.  Thus this gauge bundle is not directly visible
from the orbifold point of view.

\section{Consistent Compactifications of Super Yang--Mills theory coupled to Supergravity}
\labl{sc:consistent}

In the previous section we have constructed some gauge backgrounds on
the blowup $\cM^n$ of the orbifold $\Cplx^n/\Intr_n\,$. We have
required that they satisfy the Hermitean Yang--Mills equations on the
resolution. When these conditions are fulfilled, the background
preserves $N=1$ supersymmetry in six or four dimensions, depending 
on whether $n=2$ or $n=3\,$, respectively. In the following we will
keep $n$ generic, but have applications for these cases in mind. When
the supersymmetric gauge theory is coupled to supergravity, we
encounter one further (topological) consistency requirement. This 
condition results from the Bianchi identity of the 2--form $B$ of the
supergravity multiplet,   
\equ{
\d \cH ~=~ 
\tr \cR^2 ~-~ \tr (i\cF)^2~, 
}
where $\cH$ is its 3--form field strength. Both the trace over the
curvature 2--form $\cR \in \SU{n}$ and the $\U{16} \subset \SO{32}$
gauge background $i\cF$ are performed in fundamental representations
of $\SU{n}\,$, so no relative normalization factor is required. By
Stoke's theorem the integrated Bianchi identity over a closed 4--cycle
$C^4$ vanishes~\cite{Witten:1984dg}: 
\equ{
0 ~=~ \int_{C^4} \Big\{ \tr \cR^2 ~-~ \tr (i\cF)^2 \Big\}~.
\labl{IntdH}
}
To investigate the consequences of the integrated Bianchi identity for
the blowup $\cM^n$ of the orbifold $\Cplx^n/\Intr_n\,$, we need to
determine the 4--cycles of $\cM^n\,$.

To describe the compact and non--compact cycles of the resolution
manifold $\cM^n\,$, it is important to remember that this space was 
constructed as a cone over $\CP^{n-1}\,$. Hence, many cycles of $\cM^n$
are inherited from $\CP^{n-1}\,$, therefore we describe the relevant
cycles of this base space first: Obviously, $\CP^k$ is a $2k$--cycle
itself, in particular, $\CP^1$ is a 2--cycle and $\CP^2$ is a 4--cycle. 
Moreover, for any  group $G$ with $\gp_1(G) = 1$, we have 
$\gp_2(G/H) = \gp_1(H)$ for a proper subgroup $H \subset G\,$. 
Because $\CP^{n-1} = \SU{n}/\U{n\!-\!1}\,$, 
this implies that $\gp_2(\CP^{n-1}) = \gp_1(\U{1}) = \Intr\,$.  Since 
the homology groups can be thought of as the Abelian part of the 
fundamental groups, we conclude that $H_2(\CP^{n-1}) = \Intr\,$. This
means that $\CP^{n-1}$ has a non--contractible 2--cycle, which can 
be represented as the embedding of $\CP^1$ into $\CP^{n-1}$. 
Using these cycles of $\CP^{n-1}$ we can describe the 4--cycles of the
resolutions $\cM^2$ and $\cM^3\,$. The manifold $\cM^2$ is four
dimensional hence it is its own non--compact 4--cycle. The 
resolution $\cM^3$ (and all other $\cM^n\,$, $n>2\,$) has two
4--cycles: At the point $x=0$ the resolution $\cM^3$ looks like a
$\CP^2\,$, hence $\CP^2$ is a compact 4--cycle of $\cM^3\,$. In
addition, the space $\cM^3$ contains a real four dimensional manifold 
$\cM^2\,$, which defines a second non--compact cycle of
$\cM^3\,$. Below we discuss the resulting consequences of 
integrated Bianchi identities in six and four dimensional models.

\subsection{Consistent resolution of $\boldsymbol{\Cplx^2/\Intr_2}$ models}
\labl{sc:consistencyZ2}

The Bianchi identity integrated over the resolution $\cM^2$ becomes 
\equ{
\int_{\der \cM^2} \cH 
 ~=~ \int_{\cM^2} \Big\{ \tr \cR^2 ~-~ \tr (i\cF)^2 \Big\}~, 
\labl{IntdHnoncomp}
}
where the boundary $\der \cM^2$ at $x\ra \infty$ has the topology of
$\CP^1 \times S^1\,$. If the 3--form $\cH$ is trivial at $x \ra \infty$,
this condition reduces to the compact case, and this integral
vanishes. Making this simplifying assumption, we see that the standard
embedding is of course a solution, because the Bianchi identity is
satisfied locally. The $\SU{n\!-\!1}$ bundle~\eqref{SUn-1bundle} does
not exist on $\cM^2\,$, with $n=2\,$. For the $\U{1}$
bundles~\eqref{U1bundle} the vanishing integrated Bianchi identity
implies that   
\equ{
V^2 ~-~ 6 ~=~ 0~. 
\labl{IntBianchi6D}
} 
The relevant integrals \eqref{trR2int} and~\eqref{F2int} are evaluated
in appendix~\ref{sc:integrals}. This condition is similar to the
relation for fractional instantons given
in~\cite{Conrad:2000tk,Conrad:2000cw}. The solutions to the integrated
Bianchi conditions and the resulting gauge groups are given in
table~\ref{tb:Z2resMod}.

\tbZtworesMod

To obtain the spectra of the models as given in
table~\ref{tb:Z2resMod} we start from the anomaly polynomial $I_{12}$
of the Majorana--Weyl gaugino in ten
dimensions~\cite{Witten:1984dg,gsw_2}  
\equ{
I_{12} ~=~ \frac 12\, \frac 1{(2\pi)^5}\Big[ 
- \frac 1{720} \tr (i \fF)^6 ~+~ \frac 1{24\cdot 48} \tr(i \fF)^4 \tr\, \fR^2
~-~ \frac 1{256}\tr(i \fF)^2 \Big(
\frac 1{45} \tr\, \fR^4 + \frac 1{36} (\tr\, \fR^2)^2
\Big) 
\non \\[2ex] 
~+~ \frac {496}{64} \Big(
\frac 1{2\cdot 2835} \tr\, \fR^6 
+  \frac 1{4\cdot 1080} \tr\, \fR^2 \tr\, \fR^4
+ \frac 1{8\cdot 1296} (\tr\, \fR^2)^3
\Big)  
\Big]~. 
\labl{I12anom}
}
We then expand $\fR = \cR + R$ and $i\fF = i\cF + i F$ around the
background set by the curvature $\cR$ of the blowup $\cM^2$ and the
field strength $i\cF$ of the gauge bundle of the corresponding 
model; $R$ and $i F$ denote the curvature and gauge field strength in
six dimensions. This gives an expression for the anomaly polynomial 
\equ{
2 (2\gp)^5\, I_{12} ~=~  
- \frac {1}{256} 
\Big[ 
(i\cF)^2\, \tr \big[ H_V^2 \big] - \frac {496}{12} \tr\, \cR^2 
\Big]
\Big( 
\frac 1{45} \tr\, R^4 +  \frac 1{36} (\tr\, R^2)^2
\Big) 
\\[2ex] 
- \frac 1{48} 
\Big[ 
(i\cF)^2\, \tr \big[ H_V^2 (iF)^4 \big] - \frac 1{12} \tr\, \cR^2 \, \tr (iF)^4
\Big]
~+~ 
 \frac 1{192} 
\Big[ 
(i\cF)^2\, \tr \big[ H_V^2 (iF)^2 \big] - \frac 1{12} \tr\, \cR^2 \, \tr (iF)^2
\Big] \tr\, R^2~, 
\non
}
where $H_V = V^I H_I\,$. Integrating this expression over $\cM^2\,$,
using~\eqref{trR2int} and~\eqref{F2int} again, and some group
theoretical trace identities, the six dimensional anomaly polynomial
can be cast into the form   
\equ{
(2\gp)^3 I_{8} ~=~ 
- \frac 1{24} \tr \big[ \shalf N_V \, (iF)^4 \big] 
+ \frac 1{96} \tr \big[ \shalf N_V \, (iF)^2 \big] \, \tr\, R^2
- \frac {\tr[ \shalf N_V]}{128} 
\Big( 
\frac 1{45} \tr\, R^4 +  \frac 1{36} (\tr\, R^2)^2
\Big)~.  
}
The operator 
\equ{ 
N_V ~=~ \frac 14 \Big( H_V^2 - \frac 12 \Big) 
\labl{multiZ2}
}
counts the number of matter hyper multiplets in various
representations of the unbroken subgroup of $\SO{32}\,$. The trace
$\tr$ in this expression is taken over the full adjoint of
$\SO{32}\,$. It has to be  decomposed into irreducible representations
of the unbroken gauge group. On each irreducible representation the
operator $H_V$ and therefore $N_V$ have  definite eigenvalues. In
particular on the adjoint of the unbroken group we find $N_V =
-1/8\,,$ while on all other representations $N_V$ is positive, see
table~\ref{tb:Z2resMod}. This reflects the fact that the chiralities
of the gauginos and the matter hyperinos is opposite.

Also the values of the multiplicity factors in table~\ref{tb:Z2resMod}
can be understood easily by comparing these numbers with the spectra one
expects on the compact orbifold $T^4/\Intr_2\,$: The bulk states give
an 1/16 of the anomaly at each of the 16 fixed points of
$T^4/\Intr_2\,$. Because all the charged bulk states come from the
gauge field and the gaugino, which form doublets under the $\SU{2}$
holonomy group, we get their contributions two times, hence we obtain
a factor $1/8\,$. In this table we also encounter the factor $17/8$, which
means that the states both arise as fixed point states (four of them
at a given fixed point) and a single bulk state. Finally, the factor
$7/8$ arises from two fixed point states. These states are precisely
those needed to supply the opposite chirality of the gaugino states that
correspond to the symmetry breaking of $\SO{28} \times \SU{2} \times
\SU{2}$ to the  gauge group of the corresponding model. Again because
of $\SU{2}$ holonomy, this means that $1/8$ of the fixed states disappears
via the Higgs mechanism to form massive vector multiplets, leaving a
multiplicity factor $7/8\,$.

The procedure of integrating the anomaly polynomial
coincides with computing the Dirac indices on the resolution $\cM^2$
using Atiyah--Singer theorems. We have confirmed that the  irreducible
gauge anomalies cancel using techniques explained
in~\cite{Erler:1993zy,Berkooz:1996iz}.

\subsection{Consistent resolution of $\boldsymbol{\Cplx^3/\Intr_3}$ models}
\labl{sc:consistencyZ3}

\tbZthreeresMod

On the blowup $\cM^3$ of the orbifold $\Cplx^3/\Intr_3$ the Bianchi 
identity gives rise to two conditions, because there are two
independent 4--cycles: $\CP^2$ at the singularity and $\cM^2\,$.  As
we discussed above, the Bianchi identity integrated over $\cM^2$ can
in principle have a non--vanishing boundary integral over $\cH\,$, but
this boundary contribution vanishes, if we assume that the background
for $\cH$ is trivial. Under this assumption we have in principle two
independent consistency requirements from~\eqref{IntdH}:   
\equ{
\int_{\CP^2} \Big\{ \tr \cR^2 ~-~ \tr (i\cF)^2 \Big\}~=~ 0~, 
\qquad 
\int_{\cM^2} \Big\{ \tr \cR^2 ~-~ \tr (i\cF)^2 \Big\}~=~ 0~.
\labl{IntdHCP2cM2}
} 
Here $\cM^2$ denotes the space $\cM^3$ with, say $z^1=0\,$.
The easiest solution of these conditions is of course again the
standard embedding.

We describe the solutions of these two consistency conditions  
for the $U(1)$ gauge backgrounds~\eqref{U1bundle}. First of all,  
we find that the two conditions are equivalent. Indeed, the first
condition gives 
\equ{
-48 \gp^2 ~=~ \int_{\CP^2} \tr \cR^2 
~=~
\int_{\CP^2} \tr(i\cF_V)^2 ~=~ - 4 \gp^2\, V^2~,
}
while the second reads 
\equ{
16 \gp^2 ~=~ \int_{\cM^2} \tr \cR^2 
~=~
\int_{\cM^2} \tr(i\cF_V)^2 ~=~ \frac{4 \gp^2}3\, V^2~. 
} 
To obtain these results we have used the integrals~\eqref{trR2int}
and~\eqref{F2int}. Hence both conditions are equivalent, and imply
that the vector $V$ has to satisfy  
\equ{
V^2 ~=~ 12~. 
\labl{IntBianchiZ3}
}
The solutions to this condition and the resulting gauge groups and
spectra are collected in table~\ref{tb:Z3resMod} for the $\SO{32}$
theory.

To obtain the spectra of the models given in table~\ref{tb:Z3resMod},
we again start from the anomaly polynomial~\eqref{I12anom} for the
gaugino in ten dimensions.  Because the result for the standard
embedding are well--known, we only focus on the $U(1)$ gauge bundles
here. We insert the background of the resolution manifold $\cM^3$ and
the gauge bundle of the corresponding model into this anomaly
polynomial. Using the branching of the Lorentz and gauge group, and
some additional group theoretical properties, the anomaly polynomial
$I_{12}$ on the resolution $\cM^3$ can be written as  
\equ{
2 (2\pi)^5 I_{12} ~=~ (i\cF)^3 \Big\{ 
- \frac 1{36} \tr[ H_V^3 (iF)^3 ] 
~+~ \frac 1{9\cdot 32} \tr[ H_V^3 \, iF ] \tr\, R^2
\Big\} 
\non \\[2ex] 
~+~ 2\, i\cF\, \tr\, \cR^2 
\Big\{
\frac 1{9\cdot 32} \tr[ H_V (iF)^3 ] 
~-~ \frac 1{9\cdot 256} \tr[ H_V iF ] \tr\, R^2 
\Big\}~. 
}
This expression is integrated over $\cM^3$, using 
the expressions~\eqref{F3int} and~\eqref{trR2F1int} of
appendix~\ref{sc:integrals}, to give 
\equ{
-i\, (2\pi)^2\, I_6 ~=~ \frac 1{6} \tr \big[ \frac 12 \,N_V (iF)^3 \big]
~-~ \frac 1{48} \tr\big[ \frac 12\, N_V\, iF \big] \, \tr\, R^2~. 
\labl{I6anom} 
} 
This expression for the anomaly in four dimensions can be used to
read off the chiral spectrum of the model. The operator 
\equ{ 
N_V ~=~ \frac 16 \Big( - \frac 13 H_V^2 + 1 \Big) H_V
\label{NVZ3}
}
gives the multiplicity of the irreducible representations after
decomposing the trace $\tr$ again. It has been normalized such that
in~\eqref{I6anom} we take into account that from the adjoint of
$\SO{32}$ all complex representations appear in conjugate pairs. 
The charges $H_V$, and therefore $N_V\,$, of such a pair are opposite,
so for the four dimensional anomaly they contribute twice.

\tbsuumodels

In table~\ref{tb:Z3resMod} the charges $H_V$ and the multiplicity
values $N_V$ are  indicated. Most multiplicities are multiples of
$1/9\,$. The reason for this is that we compute the spectrum on the
resolution of the non--compact orbifold $\Cplx^3/\Intr_3\,$. As has
been shown in~\cite{Gmeiner:2002es} the anomaly of bulk fields at a
fixed point is $1/27$ of the zero mode anomaly on $T^6/\Intr_3\,$, but
each bulk state has a multiplicity of three due to the $\SU{3}$
holonomy. Hence in total we find a factor of $1/9\,$. For states
localized at the orbifold fixed point we do not have such fractional
multiplicity factors. In table~\ref{tb:Z3resMod} these states can
be spotted easily, either they have a multiplicity factor of 
$1\,$, or $26/9\,$. In the latter case $1/9$ of the fixed point state
has paired up with a bulk state.

Using the shift vectors as given in table~\ref{tb:Z3resMod} it is
straightforward to read off the gauge enhancement in the blow down
limit. In table~\ref{tb:Z3orMod} the resulting gauge groups are
displayed to facilitate the comparison in section~\ref{sc:hetZ3} of our
blowup models with the heterotic $\Intr_3$ orbifold models in four
dimensions. Here we only notice that all gauge groups except 
$\SO{32}$ of heterotic $\Intr_3$ models are recovered in this
limit. To see how this model could arise, we would like to make some  
comments on models in which the $\U{1}$ bundles~\eqref{U1bundle} and
multiple embeddings (say $k$ times) of the $\SU{2}$ gauge
background~\eqref{SUn-1bundle} are combined. As long as we make sure
that in this combined embedding all parts still commute with each
other, we are guaranteed that the Hermitian Yang--Mills equations
remain satisfied. In this case the integrated Bianchi conditions 
for $\CP^2$ and $\cM^2$ give the conditions 
\equ{ 
- 48\gp^2 ~=~ k \, 6\pi^2 ~-~ 4\pi^2\, V^2~, 
\qquad 
\int_{\der \cM^2} \cH 
 ~=~ 16\gp^2 ~-~ \frac {4\gp^2}3 \, V^2
~=~ - k\, 2\gp^2~, 
\label{condSU2U1}
} 
respectively, which are not equivalent anymore. The second equation
says that we can allow for multiple embeddings of $\SU{2}$ bundles,
only if we have non--trivial $\cH$ flux at infinity (i.e.\ at $\der
\cM^2$). But then the geometrical background needs to have torsion and
hence is  non--\Kh~\cite{Strominger:1986uh,LopesCardoso:2002hd}.
This means that the Ricci--flat \Kh\ manifold $\cM^3\,$, discussed in
this work, does not define the appropriate setting to investigate such
gauge bundle configurations.

Even though the full explicit construction of models with such
combined $\SU{2}$--$\U{1}$ bundles lies beyond the scope of
this paper, let us make some speculations: Let us assume that the
integrals in the required torsion background still lead to the first
equation~\eqref{condSU2U1}. This condition simply equates the
instanton numbers on $\CP^2$ (upto a normalization factor), so one may
expect that their values stay the same when one introduces torsion. We
see that the model with $n=4$ and $V=(0^{14},3^2)$ satisfies this
condition. This model in blow down reproduces the $\SO{32}$ model,
which we are not able to construct using the standard embedding or
$\U{1}$ bundles alone.  Using such combined $\SU{2}$--$\U{1}$ bundles,
one can consider many other models that satisfy the consistency
condition~\eqref{condSU2U1}. The resulting solutions for the $\SO{32}$
theory and  the gauge groups on the resolution and in the blow down
limit are given in table~\ref{tb:su2u1models}. (A similar table can be
produced for the $\E{8}\times\E{8}'$ case but will not be given here.)
The reason that in the blow down the rank of the gauge group is
enhanced is because the $\SU{2}$ bundles disappear inside the orbifold
singularity: they only have support on the $\CP^2\,$, that is on 
the blowup of the orbifold singularity. The blow down gauge groups are
precisely all possible gauge groups for $\Intr_3$ heterotic $\SO{32}$ 
orbifolds, except the $\SO{2}\times\U{15}$ model. Let us emphasize 
that our work does not strictly speaking prove that these models
exist, but gives strong hints that they might.

\section{Matching with Heterotic Orbifold Models} 
\labl{sc:heterotic}

We now compare our results on the resolutions $\cM^n$ of
$\Cplx^n/\Intr_n\,$, for $n=2,3$, using field theory techniques only,
with the heterotic string on such orbifolds. Before we turn to the
details of this comparison, we first review the requirements on
heterotic orbifolds, see
e.g.~\cite{dixon_85,Dixon:1986jc,ibanez_87,ibanez_88}.

In the heterotic string a perturbative $\Cplx^{n}/\Intr_n$ or
$T^{2n}/\Intr_n$ orbifold is completely specified by the action of 
the orbifold operator on the spacetime geometry and on the gauge
bundle. (We describe only the heterotic $\SO{32}$ string here, as the
$\SO{32}$ gauge theory has mostly been focused on in this work; the
extension to the $\E{8}\times\E{8}'$ case is straightforward.)
The geometric action is required to be well--defined on bosons
and spinors, and leaves one spinor invariant to preserve some
supersymmetry. The geometrical shift~\eqref{geomshift} already
satisfies all these requirements in the field theoretical description as
was discussed below that equation. The orbifold action on the gauge
bundle is instead specified by the vector $V$, as explained in
section~\ref{sc:u1bundles}. In addition to the demand that its
entries, $V^I\,$, should either all be integer or all half-integer, we
need to require that 
\equ{ 
\sum V^I ~=~  0 ~{\rm mod}~ 2~. 
}
The reason for this is that the (massive) spectrum of the heterotic
$\SO{32}$ string also contains positive chirality spinorial
representations. This condition ensures that the $\Intr_n$ orbifold
action has also order $n$ on positive chiral spinors as well. We have
enforced this constraint on all gauge shift vectors given in
tables~\ref{tb:Z2resMod},~\ref{tb:Z3resMod} and~\ref{tb:Z3resE8} by
including some appropriate minuses of some shift vector entries. For
the field theory models, we have discussed,  this constraint is
irrelevant because the models and spectra are identical. Finally, we
get to the only real string condition: Modular invariance of the
partition function imposes the following relation  
\equ{ 
V^2 ~=~ \gf^2 ~{\rm mod}~2n 
\labl{ModInv}
}
among the geometrical and gauge shifts.

The various string conditions described here are reflected in the
construction of the smooth resolutions of the orbifold singularity and
its $\U{1}$ bundles: The requirement of preserving a certain amount of 
supersymmetry forced the resolution to be Calabi--Yau, i.e.\ a 
Ricci--flat \Kh\ manifold with $\SU{n}$ holonomy. In the blow down
limit we read off a geometrical shift, see~\eqref{geomshiftID}, which
satisfies the above requirements. Similarly, integrality of the first
Chern class~\eqref{Integral1stCh} of the $\U{1}$ bundle on the blowup
was linked to the orbifold conditions of the gauge shift $V$
via~\eqref{blowdownU1}. The modular invariance
condition~\eqref{ModInv} should be identified with the integrated
Bianchi identity condition~\eqref{IntdHnoncomp} on the resolution. 
The latter condition is much more restrictive, indeed, all the gauge
shifts listed in tables~\ref{tb:Z2resMod},~\ref{tb:Z3resMod}
and~\ref{tb:Z3resE8} satisfy the corresponding modular invariance
condition~\eqref{ModInv} of $\Cplx^2/\Intr_2$ and $\Cplx^3\Intr_3$ orbifold,
respectively.

The conditions described here apply both to the non--compact orbifolds
$\Cplx^n/\Intr_n$ and to their compact relatives $T^{2n}/\Intr_n\,$. The
compact orbifolds can be equipped with discrete Wilson lines, which
need to fulfill additional consistency conditions~\cite{ibanez_87,ibanez_88}. 
These extra requirements are equivalent to local modular invariance
conditions~\eqref{ModInv} at each of the fixed points, for the local
gauge shift vectors of those fixed points, as has been demonstrated
in~\cite{Gmeiner:2002es}. The reason for these local conditions can
also be understood from the blowup perspective:  We expect, that
patching of various copies of the resolution geometry with gauge
bundles only gives mild modifications in the vicinity of the gluing
areas, leading to a satisfactory description of the whole compact
orbifold. Then the effect of discrete Wilson lines, i.e.\ different
local gauge shifts, corresponds to a space constructed by gluing
patches with the same base geometry $\cM^n$ but different $\U{1}$
bundles. Since each $\cM^3$ contains a $\CP^2$ we find an integrated
Bianchi for each of the resolved fixed points, which corresponds to
the local modular invariance conditions. In the two dimensional
complex case, the local integrated Bianchi identity on the resolution
$\cM^2$ gives the analog of the modular invariance condition provided
that we do not have non--trivial $\cH$ flux. For simplicity, 
we consider only orbifold models without discrete Wilson lines, so
that the comparison between the spectra of models on the blowups,
$\cM^2$ and $\cM^3$,  and the compact orbifolds, $T^4/\Intr_2$ and
$T^6/\Intr_3$, respectively, is clearer.

The aim of the remainder of this section is to perform comparisons at
different levels. First of all we can compare the gauge groups of the
blowups with the heterotic orbifold models. In general the gauge
groups on the resolution are smaller than the corresponding orbifold
ones. To obtain a fair comparison one should switch on some VEVs of
fields in the heterotic orbifold model to break to the groups, that
appear on the blowup. This is a tedious and difficult exercise,
because it requires a good understanding of the potential of the
model. An easier procedure is to consider the blow down limit of the
blowup models. As was explained in section~\ref{sc:u1bundles} in this
limit the $\U{1}$ bundle models can be directly reinterpreted as
non--trivial orbifold boundary conditions. This allows one to directly
read off from the shift vector $V\,$, that defines the $\U{1}$ bundle,
and what the gauge group is in the blow down limit. For models that in
this limit have the same gauge group, one can subsequently ask to what
extend also the matter spectra are identical.

The matter spectra of heterotic string on orbifolds fall into two
categories: untwisted and twisted matter. Untwisted matter are simply
those states of the original Yang--Mills supergravity theory that
survive the orbifold projections. The twisted string states are
additional states, that arise because of open strings on the covering
space of the orbifold can appear as closed strings on the orbifold
itself. Their massless excitations are localized at the orbifold fixed
points. The string theory prediction of these twisted states is rather
mysterious from the point of view of orbifold field theories. On
the concrete resolutions with bundles constructed in this work, we
would like to investigate how much of the twisted matter can be
recovered using field theory techniques only.

\subsection{$\boldsymbol{T^4/\Intr_2}$ models}
\label{sc:hetZ2}

In this subsection we compare our results on the two dimensional
complex Eguchi--Hanson space with $\U{1}$ bundles, summarized in
table~\ref{tb:Z2resMod} with heterotic orbifolds on $T^4/\Intr_2\,$. 
The discussion here can be brief, because a related study of merging 
heterotic models on this orbifold and its (unique) blowup $K3$ has
been carried out recently~\cite{Honecker:2006qz}.

The characterization of a line bundle model there can be identified
with our classification using the shift $V\,$: Each entry
$V^I$ indicates that the $V^I$--th power of the fundamental line
bundle $L$ is employed. Using this identification we confirm that the
models with $V=(0^{10},1^6)$ and $(\frac 12^{15},\sm\frac 32)$ are
reproduced identically both on the level of the gauge groups as well
as the spectra. (When comparing the spectra one should take into
account that we consider the resolution of a single fixed point of
$T^4/\Intr_2$, while in~\cite{Honecker:2006qz} the blowup of
$T^4/\Intr_2$ as a whole, i.e.\ $K3\,$, is considered, hence our
spectra have to be multiplied by a factor $16\,$.) Our $\U{1}$ bundle
model with $V=(0^{13},1^2,2)$ was not discussed
in~\cite{Honecker:2006qz}.\footnote{The reason that this model was not
discussed there is, that the aim of that paper was to find a $K3$
realization of the spectra of each of the known $T^4/\Intr_n$ orbifold
models, but not to give an exhaustive classification of all possible
$K3$ models.} We have checked that this model in blow down corresponds
to the standard embedding orbifold model. The comparison is exact on
the level of the spectrum: Because of the gauge enhancement of
$\SO{26} \times \U{1}$ to $\SO{28}$ some massive gauge fields and
gauginos in the blowup give precisely those extra hyper multiplet
states to complete the massless spectrum of the heterotic standard
embedding orbifold model with $V=(0^{14},1^2)\,$. Hence, all three
$\U{1}$ bundle models of table~\ref{tb:Z2resMod}, that satisfy the
vanishing integrated Bianchi identity, correspond directly to the
three heterotic string orbifold models in blow down. The matching of
all three models is exact including the full chiral matter spectra:
From the gaugino we were able to reconstruct both the full untwisted
and twisted string states.

\begin{figure}
\begin{center} 
\raisebox{0ex}{\scalebox{0.36}{\mbox{\input{z2.pstex_t}}}}
\end{center} 
\captn{To obtain a realization the $(L,L)$ model
of~\cite{Honecker:2006qz} as a blowup of $T^4/\Intr_2\,$ equal Wilson
lines $A$ have to be put in both directions of the first torus. As can
be seen from the local shift vectors $V$, $V+A$ and $V+2\,A\,$, the
fixed points are not treated democratically on the blowup. In the
orbifold limit these Wilson lines are irrelevant and all fixed points
are equivalent. 
} 
\labl{fg:Z2model}
\end{figure}

To close this subsection, we make a few comments on the line bundle
model $(L, L)$ found in~\cite{Honecker:2006qz}. As observed there,
this model cannot be realized in a democratic way: i.e.\ put an equal
gauge flux on each of the 16 cycles, that correspond to the fixed
points of the orbifold $T^4/\Intr_2\,$. (If one insists on doing so
one gets a shift $V = (0^{14}, \sqrt{3}{}^2)\,$, which is not allowed.) 
This model can be understood as a blowup of the orbifold
$T^4/\Intr_2$ with shift vector $V=(0^{14},\sm 1^2)$ and
equal discrete Wilson lines $A=(0^{14},2^2)\,$ in both directions on
the first torus. As is depicted in figure~\ref{fg:Z2model} this model has
fixed points with three different shift local gauge vectors: Four
fixed points have the local shift $V$ equal to the orbifold shift,
eight fixed points have the shift $V+A=(0^{14},1^2)\,$, and finally
four fixed points carry the shift $V+2\,A=(0^{14},3^2)\,$. All these
shifts satisfy the local version of the modular invariance
condition~\eqref{ModInv}. (From the orbifold perspective these
Wilson lines are irrelevant, as the local shift vectors are
equivalent in blow down.) However, all fixed points have non--vanishing
integrated Bianchi identities: 
\equ{
V^2 ~-~ 6 ~=~ (V+A)^2 ~-~ 6 ~=~ -4~,  
\qquad 
(V+2\,A)^2 ~-~ 6 ~=~ 12~,  
}
using that~\eqref{IntBianchi6D} is the condition for having a
vanishing one. This means that all the fixed points carry non--trivial
$\cH$ flux, and hence have torsion, nevertheless the total flux on the
blowup of the compact $T^4/\Intr_2$ cancels identically. As all local
gauge shifts are proportional, the gauge symmetry on the compact
blowup is: $\SO{28}\times \U{2}\,$; precisely as the model as the
$(L,L)$ model in~\cite{Honecker:2006qz}.

\subsection{$\boldsymbol{T^6/\Intr_3}$ models}
\label{sc:hetZ3}

\tbZthreeorMod

We now turn to the comparison between heterotic $\SO{32}$ models on
$T^6/\Intr_3$ and the models summarized in table~\ref{tb:Z3resMod}
with $\U{1}$ bundles on the blowup $\cM^3\,$. The classification of
heterotic $\Intr_3$ models was first given in~\cite{Giedt:2003an} and
reviewed in~\cite{Choi:2004wn,Nilles:2006np}.\footnote{Here we ignore the
heterotic orbifold model with  trivial orbifold boundary conditions,
which has $\SO{32}$ as the surviving gauge group. As noticed at the end of
subsection~\ref{sc:consistencyZ3} this model can only be recovered if
we allow for non--trivial $H$ flux. In this section we only compare
with blowup models that do not carry this type of flux.} The standard
forms of the possible shift vectors of the heterotic orbifold modes are
listed in the first column of table~\ref{tb:Z3orMod}. For most of the
rows there seems to be a mismatch between this classification of the
gauge shifts and the one given in table~\ref{tb:Z3resMod}, repeated in
the second column of table~\ref{tb:Z3orMod}. Of course we can only
really compare the orbifold shifts with the shifts, characterizing the
$\U{1}$ bundles on the blowup, after we have taken the blow down
limit. Moreover, one has to take into account that two different shift
vectors lead to fully equivalent heterotic orbifold theories, when
only some signs of their entries differ, or when their difference
equals a  vectorial or spinorial weight. With this in mind, it is not
difficult to confirm that the same gauge groups, listed in the third
column of table~\ref{tb:Z3orMod}, are obtained in the blow down limit
of the resolution models and in the heterotic orbifold models. We see
that some orbifold models can be matched with two different blowup
models. In these cases, we have shift vectors that are equivalent in
the orbifold limit, but not in the blowup regime: there they produce
models with different gauge groups, see table~\ref{tb:Z3resMod}.

The comparison between orbifold models and blowup models can 
be extended to the spectra. As discussed above, matching at this level
is best studied in the blow down limit of the smooth realizations. 
Because of the gauge enhancement in this limit the matter 
states are reorganized into representations of the enhanced gauge
groups. This regrouping of representations is encoded in the
differences in the matter spectra of table~\ref{tb:Z3resMod}
and the one but last column of table~\ref{tb:Z3orMod}. All states
needed to form the bigger representations of the enhanced gauge group
are already present in the spectra of table~\ref{tb:Z3resMod}. (The
only exception to this is the state $(\rep{8}_+,\rep{1})$ in the one
but last row of table~\ref{tb:Z3orMod}: It is obtained by combining
the $(\crep{6},\rep{1})$ state of the $(\frac 12{}^{12},\frac 32{}^4)$
model of table~\ref{tb:Z3resMod} with a non--chiral pair of singlets
w.r.t.\ the blowup gauge group.) We wish to  stress, that the spectra 
in the one but last column is obtained from the ten dimensional
gaugino alone. When comparing these spectra to those of heterotic
$\Intr_3$ orbifold~\cite{Choi:2004wn}, we see that only the states in
the final column of table~\ref{tb:Z3orMod} are not reconstructed from
the ten dimensional gaugino states. We see that exact matching occurs
in three models. For two other models the  matching is exact up to a
single missing singlet on the blow down side. In the two remaining
models the mismatch is slightly larger: In addition to a single
singlet, also one time the vector $(\rep{26},\rep{1})$ and one time
the spinor $(\rep{8}_+,\rep{1})$ failed to appear from the gaugino on
the resolution. However, non of these state are chiral. Therefore, to
summarize, we can say, that the matching is always extact at the level
of the {\it chiral} spectrum, thus the mismatch can be due to states
that can easily get a mass.

The analysis presented above can be repeated for the heterotic
$\E{8}\times\E{8}'$ super Yang--Mills theory in ten dimensions. We
will not dwell on the details here, but for completeness, we have also
determined the $\E{8}\times\E{8}'$ $T^6/\Intr_3$ models, by
identifying the gauge shifts that satisfy the integrated Bianchi
identity~\eqref{IntBianchiZ3} (upto interchanges of the two $\E{8}$
factors), see table~\ref{tb:Z3resE8}.  
This table shows that there are only eight possible resulting gauge
groups on the resolution. Because of gauge enhancement in the blow
down limit we have only four possible gauge groups. However, even
though many shift vectors give rise to the same gauge group on the
resolution, this does not necessarily mean that the corresponding
models are equivalent, because their spectra can be different. To
establish that some of the models are identical, one has to confirm
that on all matter representations the operator $N_V\,$, defined
in~\eqref{NVZ3}, gives the same multiplicities. As for the ten
dimensional $\SO{32}$ theory also in the $\E{8}\times\E{8}'$ case, we
do not recover the models with trivial gauge embeddings, even 
though they arise as heterotic string
models~\cite{Gmeiner:2002es,Choi:2004wn}. 
The speculations at the end of subsection~\ref{sc:consistencyZ3}, that
this model could arise by combining $\SU{2}$ and $\U{1}$ bundles with
torsion on the resolution, can be extended to  the $\E{8}\times\E{8}'$ theory.

\section{Conclusions}
\labl{sc:concl}

\tbZthreeresE

We have described blowups of $\Cplx^n/\Intr_n$ orbifolds as complex
line bundles over $\CP^{n-1}\,$. Our parameterizations of the metrics
of these Eguchi--Hanson spaces are uniquely determined by demanding
that they possess an $\SU{n}$ symmetry as the original orbifolds
do. Technically this is achieved by reducing the problem of finding
Ricci--flat \Kh\ manifolds to solving an ordinary differential
equation for the \Kh\ potential. This is possible because the $\SU{n}$
symmetry implies that the \Kh\ potential is a function of a single
$\SU{n}$ invariant variable. The only parameter of the resolution can
be interpreted as the volume of the $\CP^{n-1}$ located at the
resolved orbifold singularity. Once the \Kh\ potential has been
determined, it is straightforward to compute the resulting metric,
spin--connection 1--form and the curvature 2--form. The behavior of
the curvature is as expected: Away from the --would be-- orbifold
singularity it tends to zero in the blow down limit, while at the
resolved orbifold singularity the curvature explodes. In
this way it mimics the properties of a regularized delta function: it
is smooth, but becomes strongly peaked at a single point in a specific
limit, while its integral stays finite in the blow down limit.

We have constructed some examples of gauge bundles over these 
resolutions. As a cross check we directly confirmed that the
standard embedding indeed solves the Hermitian Yang--Mills equations. 
To construct $\U{1}$ gauge bundles explicitly we followed a similar
strategy as in the construction of the geometric resolution of
$\Cplx^n/\Intr_n$ itself: We insisted on the $\SU{n}$ symmetry to
guarantee, that the gauge background is determines by a single
function of the $\SU{n}$ invariant variable. The Hermitian Yang--Mills
equations then also turn into a single ordinary differential equation,
which is readily solved. Regularity determined the $\U{1}$ bundle
uniquely up to an overall normalization. This normalization is related to 
the $\Intr_n$ orbifold boundary conditions of the gauge fields in blow
down using the Hosotani mechanism. This allows us to read off the
orbifold gauge shift vector from the gauge background. In addition, we
identified an $\SU{n\!-\! 1}$ bundle over $\CP^{n-1}\,$, which trivially
also solves the Hermitian Yang--Mills equations on the 
blowup of $\Cplx^n/\Intr_n\,$. Contrarily to the standard embedding and
the $\U{1}$ bundles, this gauge background cannot be interpreted as
orbifold boundary conditions for gauge fields in the blow down phase.

We considered the ten dimensional $\SO{32}$ super Yang--Mills theory
coupled to supergravity on these backgrounds. It is well--known that
the integrated Bianchi identity for the anti--symmetric tensor of
supergravity leads to a stringent consistency condition. This
constraint is similar to the modular invariance condition of heterotic
string model building. We confirmed this explicitly by determining
all possible models on the blowups of $\Cplx^2/\Intr_2$ and
$\Cplx^3/\Intr_3$ with $\U{1}$ gauge bundles: We found only three and
seven possible models for these four and six dimensional resolutions, 
respectively. Using the procedure to determine the orbifold gauge
shift vector, we asserted, that in the blow down limit these 
models correspond to heterotic $\SO{32}$ models on $\Cplx^2/\Intr_2$
and $\Cplx^3/\Intr_3\,$. Only the heterotic $\Intr_3$ model with gauge
group $\SO{32}$ cannot be reconstructed in blowup using our $\U{1}$
bundles, as it does not satisfy the consistency condition resulting
from the integrated Bianchi identity: Contrary to modular invariance
conditions for heterotic orbifolds, it is not a condition modulo some
multiple of integers. We have conjectured, that this missing model can
be obtained in blowup when one combines $\U{1}$ and $\SU{2}$
bundles. But since the integrated Bianchi identity then implied that
there must be torsion, this background is beyond the scope of this
paper.

We have investigated whether the spectrum of the corresponding
heterotic orbifold models can be recovered in the orbifold limit. For
this it is not sufficient to merely identify the gauge shift vector: 
The full chiral charged spectrum needs to be analyzed. On a resolution
the matter states arise from the gauge field and gaugino only, while
in heterotic orbifold models also twisted string states are present.
Therefore, we computed the charged chiral spectra on the
resolutions. To do so we started from the anomaly polynomial of the
gaugino and integrated it over the resolutions of $\Cplx^2/\Intr_2$
and $\Cplx^3/\Intr_3\,$. This gives rise to anomaly polynomials for
six and four dimensions, respectively, from which the charged spectrum
can be read off easily. We found, that the spectra of the blowups and
the heterotic orbifold models match identically for $\Cplx^2/\Intr_2\,$. 
On $\Cplx^3/\Intr_3$ the spectra were not always identical, but
discrepancies are surprisingly minor: In most cases only a singlet was
missing. Moreover, it seems always to be possible to give mass to the
states that cause the mismatch by using an anomalous $\U{1}$ at 
one--loop.

While there is a clean matching between the $\U{1}$ bundle blowup
models and the heterotic orbifold models on $\Cplx^3/\Intr_3\,$, the
situation in the type--I setting is very different.  This is
interesting in the light of the assumed $S$--duality between type--I
and heterotic string~\cite{Polchinski:1995df} in the four dimensional
setting~\cite{Kakushadze:1997wx,Kakushadze:1998cd}. (To avoid
additional complications of $D5$ branes we only consider
$\Cplx^3,T^6/\Intr_3$ orbifolds here only.) For $\Cplx^3/\Intr_3$
orbifold there is just one single type--I model
known~\cite{Angelantonj:1996uy,Aldazabal:1997wi}, while we have 
obtained seven $\U{1}$ bundle resolution models from the 
ten dimensional $\SO{32}$ gauge theory, which give rise to five
different models in blow down. This conclusion is reached assuming 
that type--IIB orientifolds with D9 branes on the resolution of 
$\Cplx^3/\Intr_3$ are not crucially different from that on ten
dimensional Minkowski space. Hence, the mismatch between orbifold 
type--I and heterotic string models does not seem to signal a
complication of $S$--duality, but rather a problem of type--I 
model building itself. The type--I $T^6/\Intr_3$ orbifold model has 
untwisted charged matter only, nevertheless its spectrum has no
irreducible anomalies. The reducible anomalies are canceled by a 
Green--Schwarz mechanism that involves twisted RR--scalars, that live
at the orbifold fixed point only. This is very different from all the
models on the blowup of $\Cplx^3/\Intr_3\,$: There it is always the
bulk anti--symmetric tensor, part of the supergravity multiplet, that
cancels the reducible anomalies. Hence, one expects that this bulk
state remains the  Green--Schwarz field in the blow down limit. This
is presumably related to the different properties of anomalous $\U{1}$s
in the type--I and heterotic models, as was pointed out
in~\cite{Lalak:1999bk}. Therefore it is an interesting problem to
understand how in orbifold type--I model building the other resolution
models in blow down can be recovered.

There are various other directions in which this work can be
extended. First of all the explicit resolutions and gauge bundles
discussed in this work correspond to a very restricted class of
$\Cplx^n/\Intr_n$ orbifolds. It would be very interesting and useful
to find similar explicit resolutions of $\Cplx^2/\Intr_n$ and
$\Cplx^3/\Intr_n$ for general $n\,$. The resolutions that we discussed
in the work possess the large $\SU{n}$ rotational symmetry, therefore,
one can wonder if one can consider deformations of them that preserve
less rotational symmetry, but nevertheless reduce to the same
orbifolds in the blow down limit. The investigation of deformations
becomes even more involved when one also takes deformations of the
gauge bundles into account. Moreover, even before considering
deformations of our blowups of $\Cplx^n/\Intr_n\,$, our discussion of
their gauge bundles was limited: We have only given a number of
examples of them. In a more complete analysis one would be looking for
a full classification and explicit construction of all possible
bundles. With all of them in hand one can complete the analysis of
possible blowup models of heterotic orbifold models. This would give a
better insight into the moduli space of the heterotic
string. Moreover, we have only restricted our attention to
perturbative heterotic string vacua for simplicity. It would be
interesting to extend our analysis to non--perturbative heterotic
vacua described in~\cite{Aldazabal:1997wi}. And as we alluded to at the
end of section~\ref{sc:consistent}, we expect that a blowup with
torsion, on which we have a combination of $\SU{2}$ and $\U{1}$
bundles, could constitute the blowup of the heterotic $\SO{32}$
$\Intr_3$ model. It would be interesting to construct this blowup with
torsion explicitly, and analyze what other models we can construct in
this way.

\section*{Note added in proof}

After this work was completed we became aware of~\cite{Ganor:2002ae}
where the same gravitational and gauge backgrounds were discussed in
the context of a particular $\Intr_3$ heterotic $\E{8}\times\E{8}'$
model in strong coupling.

\section*{Acknowledgments}

We would like to thank J.\ Conrad, W.\ Israel, J.\ Stienstra and M.\ Vonk 
for useful discussions and suggestions at the initial stages of this
project. We are grateful to H.P.\ Nilles for the stimulating
atmosphere in his group where the foundations for this project were
laid and useful comments at its completion. We would like to thank G.\
Honecker and M.\ Olechowski for discussions and comments, and careful
reading of the manuscript. We thank F.\ Pl\"oger and P.\ Vaudrevange
for pointing out some typos. 
This work was partially supported by the European Union 6th framework
program MRTN-CT-2004-503069 "Quest for unification",
MRTN-CT-2004-005104 "Forces Universe", MRTN-CT-2006-035863
"Universe Net", SFB-Transregio 33 "The Dark Universe" and HE 3236/3-1
by Deutsche Forschungsgemeinschaft (DFG).

\appendix
\def\theequation{\thesection.\arabic{equation}} 

\section{Forms on $\boldsymbol{\CP^{n-1}}$ and its Line Bundle}
\labl{sc:CPn}
\setcounter{equation}{0}

In this appendix we collect useful properties of the vielbein and
$\U{n}$ connection 1--forms of  $\CP^{n-1}$, as defined
in~\eqref{CPnvielb} and~\eqref{CPnUnconn} of the main text.  
The $\SU{n}$ isometry group structure provides a useful tool to
investigate the geometry of $\CP^{n-1}$. Consider the mapping of the
$\CP^{n-1}$ to the group $\U{n}$ given by the group element 
\equ{
U ~=~ \pmtrx{
\tgch^{-\frac 12}  & i\, \gch^{-\frac 12} \, z
\\[1ex] 
i\, \gch^{-\frac 12} \, \bz &  \gch^{-\frac 12} 
}~, 
\labl{matrixU}
}
where $\gch$ and $\tgch$ are functions of the coordinates $z$ and
$\bz$, defined in \eqref{KahlOrbi} and \eqref{CPnvielb}, respectively.
It is not hard to check that $U$ is indeed an element of $\U{n}$, 
with $U^\dag U = \Id_{n}$. The Maurer--Cartan 1--form of  this 
coset is defined as $U\inv \d U$ and takes the form 
\equ{
U\inv \d \, U ~=~  i \pmtrx{ \tilde\cB & e \\ \bar e & \cB } = 
\pmtrx{
\tgch^{-\frac 12} \bder (\tgch^{\frac 12}) 
- \der (\tgch^{\frac 12}) \tgch^{-\frac 12} 
&  
i\,  \gch^{-\frac 12} \tgch^{-\frac 12}\,  \d z
\\[2ex]
i\, \d \bz \,  \tgch^{-\frac 12}\gch^{-\frac 12} 
& 
\gch^{-\frac 12} (\der - \bder) (\gch^{\frac 12})
}~.
\labl{MaurerCartan}
}
The 1--forms $e$ and $\bee$ constitute the vielbeins of $\CP^{n-1}$, 
i.e.\ $\d s^2_{\CP^{n-1}} = \bee \otimes e$. 
Using the Maurer--Cartan structure,  
\(
\d( U\inv \d U) = - (U\inv \d U)^2\, ,
\)
we obtain the following matrix identity:
\equ{
i\, \d \pmtrx{ \tilde \cB & e \\ \bar e & \cB } 
~=~ 
\pmtrx{ \tilde \cB^2 + e \, \bee & \tB \, e + e\, \cB 
\\[2ex]
\cB\,  \bee + \bee \, \tilde \cB & \bee\, e }~.
}
This thus gives a set of useful relations of how to simplify
expressions of exterior derivatives on these forms. These relations
can be used to obtain the spin connection and the curvature of
$\CP^{n-1}$ in an elegant way. Using their standard definitions we find
\equ{
\gO_{\CP} ~=~  i \tilde \cB ~- ~i \cB~, 
\qquad 
\cR_{\CP} ~=~  e \, \bee ~-~ \bee\, e~. 
} 
Notice that the trace of neither the spin connection nor the curvature
vanishes. The fact that the trace of the  curvature does not vanish 
reflects the fact that $\CP^{n-1}$ is not Ricci--flat.

In addition to these forms, we have also encountered the line bundle
1--form $\ge$ given in~\eqref{CPnvielb}. Applying an exterior
derivative on it gives 
\equ{
\d \, \ge ~=~ n \big( y \, \bee e - i \cB\, \ge \big)~, 
\qquad 
\d \, \bge ~=~ -n \big( \byy \, \bee e - i \cB\, \bge \big)~. 
} 
The exterior derivative of $X$ is given by 
\equ{
\d\,  X ~=~ \byy \ge + \bge y~. 
}

\section{Integrals over $\boldsymbol{\CP^{n-1}}$ and $\boldsymbol{\cM^n}$}
\labl{sc:integrals} 
\setcounter{equation}{0}

In this appendix we collect various integrals, that we encounter in
the main part of the text. We first give the basic integrals, next we
give various traces over powers of the curvature and gauge field
strength 2--forms, and we compute various integrals over these
expressions.

First of all the angular integrals over $\der C$ take the form 
\equ{
\frac 1{2\gp i}\int_{\der C} \ge ~=~ 
\frac {-1}{2\gp i} \int_{\der C} \bge ~=~ |y|~, 
}
where $|y|$ is taken to be constant. 
The integrals over $\CP^1$ and $\CP^2$ are given by 
\equ{
\frac 1{~ 2\gp i}
\int_{\CP^1} \bee e ~=
\frac{1}{(2\gp i)^2} 
\int_{\CP^2} (\bee e)^2 ~=~ 1~.
\labl{CPnInts}
}
Furthermore, we need the integrals 
\equ{
\frac 1{2\gp i}
\int_\Cplx \frac{\bge \ge}{(r+X)^p} ~=~ 
 \frac {1}{(2\gp i)^2} 
\int_{\cM^2} \frac{\bee e\, \bge \ge}{(r+X)^p} 
~=~ 
\frac 1 {(2\gp i)^3}
\int_{\cM^3} \frac{(\bee e)^2 \, \bge \ge}{(r+X)^p} ~=~ 
 \frac {1}{p-1} \, \frac 1{r^{p-1}}~, 
}
which only converge if $p > 1\,$.
The curvature 2--form~\eqref{Curv2expl} is an element of the algebra
of $\SU{n}\,$, which means that $\tr\, \cR = 0\,$. The traces of the
second and the third power of the curvature read 
\equ{
\tr\, \cR^2 ~=~ 
(n+1) \Big(\frac {r}{r+X}\Big)^2 
\Big[ 
n (\bee e)^2 ~-~ \frac 2{r+X}\, \bee e \, \bge \ge 
\Big]~, 
\labl{trR2}
}\equ{
\tr\, \cR^3 ~=~ 
 (1-n^2)  \Big(\frac {r}{r+X}\Big)^3 
\Big[ 
- n (\bee e)^3 ~+~  \frac {3}{r+X} (\bee e)^2 \, \bge \ge
\Big]~. 
\labl{trR3}
}
The integrals of the trace $\tr\, \cR^2$ over $\CP^{2}$ and 
$\cM^2$ can be expressed as follows 
\equ{ 
\frac 1{(2\gp i)^2} \int_{\CP^2} \tr\, \cR^2 
~=~ 
n(n+1) \Big( \frac r{r+X} \Big)^2~, 
\qquad 
\frac 1{(2\gp i)^2} \int_{\cM^2} \tr\, \cR^2 
~=~ 
- (n+1)~. 
\labl{trR2int}
} 
The integral of $\tr\, \cR^3$ over $\cM^3$ reads 
\equ{
\frac 1{(2\gp i)^3} \int_{\cM^3} \tr\, \cR^3 
~=~ 1-n^2~.
\labl{trR3int}
}
Next we consider integrals over the field strength
2--forms~\eqref{FU1basis} and~\eqref{SUn-1bundle} of the 
background $\U{1}$ and $\SU{n\!-\!1}\,$, respectively. For the $\U{1}$
bundle we have 
\equ{
\frac 1{2\gp i} \int_{\CP^1} i \cF 
~=~ 
\frac 1{\Big( 1 + \frac 1r X\Big)^{1-\frac 1n}}~, 
\qquad 
\frac 1{2\gp i} \int_{\Cplx} i \cF 
~=~ \frac 1n~, 
\labl{trF1int}
}
\equ{
\frac 1{(2\gp i)^2} \int_{\CP^2} (i \cF)^2 
~=~ 
\frac 1{\Big( 1 + \frac 1r X\Big)^{2-\frac 2n}}~, 
\qquad 
\frac 1{(2\gp i)^2} \int_{\cM^2} (i \cF)^2 
~=~ - \frac 1n~, 
\labl{F2int}
}
\equ{
\frac 1{(2\gp i)^3} \int_{\cM^3} (i \cF)^3 
~=~ - \frac 1n~. 
\labl{F3int}
}
The trace of the $\SU{n\!-\!1}$ gauge background squared and its
integral over $\CP^2$ are given by 
\equ{
\tr (i \tilde \cF)^2 ~=~ - \frac n{n-1} \, (\bee e)^2~, 
\qquad 
\frac 1{(2\gp i)^2} \int_{\CP^2} \tr (i \tilde \cF)^2 
~=~ 
- \frac n{n-1}~,  
} 
while over $\cM^2$ this integral vanishes, because $i \tilde \cF$ does
not contain the 1--forms $\ge$ and $\bge\,$.
Finally, we can consider integrals over $\cM^3$ over 6--forms that mix
both $\U{1}$ and curvature or $\SU{n\!-\!1}$ gauge field
strength. These integrals read: 
\equ{
\frac 1{(2\gp i)^3} \int_{\cM^3} i \cF \tr\, \cR^2 
~=~ - (n+1)~,
\qquad 
\frac 1{(2\gp i)^3} \int_{\cM^3} i \cF \tr (i\tilde \cF)^2 
~=~ \frac 1{n-1}~.
\labl{trR2F1int} 
}

\bibliographystyle{paper}
{\small
\bibliography{paper}

\providecommand{\href}[2]{#2}\begingroup\raggedright\begin{thebibliography}{10}

\bibitem{Candelas:1985en}
P.~Candelas, G.~T. Horowitz, A.~Strominger, and E.~Witten ``Vacuum
  configurations for superstrings'' {\em Nucl. Phys.} {\bf B258} (1985)
46--74.

\bibitem{Witten:1985bz}
E.~Witten ``New issues in manifolds of {SU(3)} holonomy'' {\em Nucl. Phys.}
  {\bf B268} (1986)
79.

\bibitem{Braun:2005ux}
V.~Braun, Y.-H. He, B.~A. Ovrut, and T.~Pantev ``A heterotic standard model''
  {\em Phys. Lett.} {\bf B618} (2005) 252--258
\href{http://www.arXiv.org/abs/hep-th/0501070}{[{\tt hep-th/0501070}]}.

\bibitem{Braun:2005bw}
V.~Braun, Y.-H. He, B.~A. Ovrut, and T.~Pantev ``A standard model from the
  {E(8) x E(8)} heterotic superstring'' {\em JHEP} {\bf 06} (2005) 039
\href{http://www.arXiv.org/abs/hep-th/0502155}{[{\tt hep-th/0502155}]}.

\bibitem{Honecker:2006dt}
G.~Honecker ``Massive {U(1)}s and heterotic five-branes on {K3}'' {\em Nucl.
  Phys.} {\bf B748} (2006) 126--148
\href{http://www.arXiv.org/abs/hep-th/0602101}{[{\tt hep-th/0602101}]}.

\bibitem{Honecker:2006qz}
G.~Honecker and M.~Trapletti ``Merging heterotic orbifolds and {K3}
  compactifications with line bundles'' {\em JHEP} {\bf 01} (2007) 051
\href{http://www.arXiv.org/abs/hep-th/0612030}{[{\tt hep-th/0612030}]}.

\bibitem{Andreas:2004ja}
B.~Andreas and D.~Hernandez~Ruiperez ``U(n) vector bundles on calabi-yau
  threefolds for string theory compactifications'' {\em Adv. Theor. Math.
  Phys.} {\bf 9} (2005) 253--284
\href{http://www.arXiv.org/abs/hep-th/0410170}{[{\tt hep-th/0410170}]}.

\bibitem{Blumenhagen:2005ga}
R.~Blumenhagen, G.~Honecker, and T.~Weigand ``Loop-corrected compactifications
  of the heterotic string with line bundles'' {\em JHEP} {\bf 06} (2005) 020
\href{http://www.arXiv.org/abs/hep-th/0504232}{[{\tt hep-th/0504232}]}.

\bibitem{Blumenhagen:2005pm}
R.~Blumenhagen, G.~Honecker, and T.~Weigand ``Supersymmetric (non-)abelian
  bundles in the type {I} and {SO(32)} heterotic string'' {\em JHEP} {\bf 08}
  (2005) 009
\href{http://www.arXiv.org/abs/hep-th/0507041}{[{\tt hep-th/0507041}]}.

\bibitem{Weigand:2005ng}
T.~Weigand ``Heterotic vacua from general (non-) {A}belian bundles'' {\em
  Fortsch. Phys.} {\bf 54} (2006) 505--513
\href{http://www.arXiv.org/abs/hep-th/0512191}{[{\tt hep-th/0512191}]}.

\bibitem{dixon_85}
L.~Dixon, J.~A. Harvey, C.~Vafa, and E.~Witten ``Strings on orbifolds'' {\em
  Nucl. Phys.} {\bf B261} (1985)
678--686.

\bibitem{Dixon:1986jc}
L.~J. Dixon, J.~A. Harvey, C.~Vafa, and E.~Witten ``Strings on orbifolds. 2''
  {\em Nucl. Phys.} {\bf B274} (1986)
285--314.

\bibitem{Ibanez:1988pj}
L.~E. Ibanez, J.~Mas, H.-P. Nilles, and F.~Quevedo ``Heterotic strings in
  symmetric and asymmetric orbifold backgrounds'' {\em Nucl. Phys.} {\bf B301}
  (1988)
157.

\bibitem{Casas:1989wu}
J.~A. Casas, M.~Mondragon, and C.~Munoz ``Reducing the number of candidates to
  standard model in the {Z(3)} orbifold'' {\em Phys. Lett.} {\bf B230} (1989)
63.

\bibitem{Katsuki:1989kd}
Y.~Katsuki, Y.~Kawamura, T.~Kobayashi, N.~Ohtsubo, and K.~Tanioka ``Gauge
  groups of {$Z(N)$} orbifold models'' {\em Prog. Theor. Phys.} {\bf 82} (1989)
171.

\bibitem{Katsuki:1990bf}
Y.~Katsuki {\em et al.} ``{$Z(N)$} orbifold models'' {\em Nucl. Phys.} {\bf
  B341} (1990)
611--640.

\bibitem{Kobayashi:1991mi}
T.~Kobayashi and N.~Ohtsubo ``Analysis on the {W}ilson lines of {Z(N)} orbifold
  models'' {\em Phys. Lett.} {\bf B257} (1991)
56--62.

\bibitem{Kobayashi:1994rp}
T.~Kobayashi and N.~Ohtsubo ``Geometrical aspects of {Z(N)} orbifold
  phenomenology'' {\em Int. J. Mod. Phys.} {\bf A9} (1994)
87--126.

\bibitem{Kawamura:1996zu}
Y.~Kawamura and T.~Kobayashi ``Flat directions in {Z}(2n) orbifold models''
  {\em Nucl. Phys.} {\bf B481} (1996) 539--576
\href{http://www.arXiv.org/abs/hep-th/9606189}{[{\tt hep-th/9606189}]}.

\bibitem{Giedt:2003an}
J.~Giedt ``{Z}(3) orbifolds of the {SO(32)} heterotic string: 1 {W}ilson line
  embeddings'' {\em Nucl. Phys.} {\bf B671} (2003) 133--147
\href{http://www.arXiv.org/abs/hep-th/0301232}{[{\tt hep-th/0301232}]}.

\bibitem{Choi:2004wn}
K.-S. Choi, S.~Groot~Nibbelink, and M.~Trapletti ``Heterotic {SO(32)} model
  building in four dimensions'' {\em JHEP} {\bf 12} (2004) 063
\href{http://www.arXiv.org/abs/hep-th/0410232}{[{\tt hep-th/0410232}]}.

\bibitem{Nilles:2006np}
H.~P. Nilles, S.~Ramos-Sanchez, P.~K.~S. Vaudrevange, and A.~Wingerter
  ``Exploring the {SO(32)} heterotic string'' {\em JHEP} {\bf 04} (2006) 050
\href{http://www.arXiv.org/abs/hep-th/0603086}{[{\tt hep-th/0603086}]}.

\bibitem{Gmeiner:2002es}
F.~Gmeiner, S.~Groot~Nibbelink, H.~P. Nilles, M.~Olechowski, and M.~Walter
  ``Localized anomalies in heterotic orbifolds'' {\em Nucl. Phys.} {\bf B648}
  (2003) 35--68
\href{http://www.arXiv.org/abs/hep-th/0208146}{[{\tt hep-th/0208146}]}.

\bibitem{GrootNibbelink:2003gb}
S.~Groot~Nibbelink, H.~P. Nilles, M.~Olechowski, and M.~G.~A. Walter
  ``Localized tadpoles of anomalous heterotic {U(1)}'s'' {\em Nucl. Phys.} {\bf
  B665} (2003) 236--272
\href{http://www.arXiv.org/abs/hep-th/0303101}{[{\tt hep-th/0303101}]}.

\bibitem{Nibbelink:2003rc}
S.~Groot~Nibbelink, M.~Hillenbach, T.~Kobayashi, and M.~G.~A. Walter
  ``Localization of heterotic anomalies on various hyper surfaces of
  {T(6)/Z(4)}'' {\em Phys. Rev.} {\bf D69} (2004) 046001
\href{http://www.arXiv.org/abs/hep-th/0308076}{[{\tt hep-th/0308076}]}.

\bibitem{Lust:2006zh}
D.~Lust, S.~Reffert, E.~Scheidegger, and S.~Stieberger ``Resolved toroidal
  orbifolds and their orientifolds''
\href{http://www.arXiv.org/abs/hep-th/0609014}{[{\tt hep-th/0609014}]}.

\bibitem{Eguchi:1978xp}
T.~Eguchi and A.~J. Hanson ``Asymptotically flat selfdual solutions to
  {E}uclidean gravity'' {\em Phys. Lett.} {\bf B74} (1978)
249.

\bibitem{pol_2}
J.~Polchinski {\em String theory vol. 2: Superstring theory and beyond}.
\newblock Cambridge, Uk: Univ. Pr. 531 P. (Cambridge Monographs On Mathematical
  Physics) 1998.

\bibitem{Joyce:2000}
D.~D. Joyce {\em Compact manifolds with special holonomy}.
\newblock Oxford University Press, 436 P. (Oxford Mathematical Monographs)
  2000.

\bibitem{Cvetic:2001zb}
M.~Cvetic, G.~W. Gibbons, H.~Lu, and C.~N. Pope ``{H}yper-{K}aehler {C}alabi
  metrics, {L**2} harmonic forms, resolved {M2}-branes, and {AdS(4)/CFT(3)}
  correspondence'' {\em Nucl. Phys.} {\bf B617} (2001) 151--197
\href{http://www.arXiv.org/abs/hep-th/0102185}{[{\tt hep-th/0102185}]}.

\bibitem{Candelas:1989js}
P.~Candelas and X.~C. de~la Ossa ``Comments on conifolds'' {\em Nucl. Phys.}
  {\bf B342} (1990)
246--268.

\bibitem{PandoZayas:2000sq}
L.~A. Pando~Zayas and A.~A. Tseytlin ``3-branes on resolved conifold'' {\em
  JHEP} {\bf 11} (2000) 028
\href{http://www.arXiv.org/abs/hep-th/0010088}{[{\tt hep-th/0010088}]}.

\bibitem{Page:1985bq}
D.~N. Page and C.~N. Pope ``Inhomogeneous einstein metrics on complex line
  bundles'' {\em Class. Quant. Grav.} {\bf 4} (1987)
213.

\bibitem{Berard-Bergery:1986}
L.~Berard-Bergery ``Quelques exemples de varietes riemanniennes completes non
  compactes a courbure de ricci positive,'' {\em C.R. Acad. Sci., Paris, Ser}
  {\bf I302} (1986) 159.

\bibitem{Calabi:1979}
E.~Calabi ``M\' etriques {K}aehl\' eriennes et fibr\' es holomorphes'' {\em
  Ann. Scient. ´Ecole Norm. Sup.} {\bf 12} (1979) 269.

\bibitem{Higashijima:2002px}
K.~Higashijima, T.~Kimura, and M.~Nitta ``{C}alabi-{Y}au manifolds of
  cohomogeneity one as complex line bundles'' {\em Nucl. Phys.} {\bf B645}
  (2002) 438--456
\href{http://www.arXiv.org/abs/hep-th/0202064}{[{\tt hep-th/0202064}]}.

\bibitem{Higashijima:2001fp}
K.~Higashijima, T.~Kimura, and M.~Nitta ``Gauge theoretical construction of
  non-compact {C}alabi-{Y}au manifolds'' {\em Annals Phys.} {\bf 296} (2002)
  347--370
\href{http://www.arXiv.org/abs/hep-th/0110216}{[{\tt hep-th/0110216}]}.

\bibitem{Higashijima:2001vk}
K.~Higashijima, T.~Kimura, and M.~Nitta ``{R}icci-flat {K}aehler manifolds from
  supersymmetric gauge theories'' {\em Nucl. Phys.} {\bf B623} (2002) 133--149
\href{http://www.arXiv.org/abs/hep-th/0108084}{[{\tt hep-th/0108084}]}.

\bibitem{Serone:2004yn}
M.~Serone and A.~Wulzer ``Orbifold resolutions and fermion localization'' {\em
  Class. Quant. Grav.} {\bf 22} (2005) 4621--4650
\href{http://www.arXiv.org/abs/hep-th/0409229}{[{\tt hep-th/0409229}]}.

\bibitem{Wulzer:2005ps}
A.~Wulzer ``Orbifold resolutions with general profile'' {\em Class. Quant.
  Grav.} {\bf 23} (2006) 1217--1240
\href{http://www.arXiv.org/abs/hep-th/0506210}{[{\tt hep-th/0506210}]}.

\bibitem{Klerk:Master2002}
G.~W.~A. de~Klerk ``The {M}c{K}ay correspondence for finite {A}belian subgroups
  of {$SL(3,\Cplx)$}'' Master's thesis University Utrecht 2002.

\bibitem{Bando:1984ab}
M.~Bando, T.~Kuramoto, T.~Maskawa, and S.~Uehara ``Structure of nonlinear
  realization in supersymmetric theories'' {\em Phys. Lett.} {\bf B138} (1984)
94.

\bibitem{GrootNibbelink:2000gq}
S.~Groot~Nibbelink ``Supersymmetric non-linear unification in particle physics:
  Kaehler manifolds, bundles for matter representations and anomaly
  cancellation''. PhD Thesis Free University Amsterdam, 2000.

\bibitem{GrootNibbelink:2000hu}
S.~Groot~Nibbelink, T.~S. Nyawelo, and J.~W. van Holten ``Construction and
  analysis of anomaly free supersymmetric {SO(2N)/U(N)} sigma-models'' {\em
  Nucl. Phys.} {\bf B594} (2001) 441--476
\href{http://www.arXiv.org/abs/hep-th/0008097}{[{\tt hep-th/0008097}]}.

\bibitem{GrootNibbelink:1999un}
S.~Groot~Nibbelink ``Line bundles in supersymmetric coset models'' {\em Phys.
  Lett.} {\bf B473} (2000) 258--263
\href{http://www.arXiv.org/abs/hep-th/9910075}{[{\tt hep-th/9910075}]}.

\bibitem{Nakahara:1990th}
M.~Nakahara ``Geometry, topology and physics''. Bristol, UK: Hilger (1990) 505
  p. (Graduate student series in physics).

\bibitem{Witten:1984dg}
E.~Witten ``Some properties of {O}(32) superstrings'' {\em Phys. Lett.} {\bf
  B149} (1984)
351--356.

\bibitem{Conrad:2000tk}
J.~O. Conrad ``On fractional instanton numbers in six dimensional heterotic
  {$\E{8} \times \E{8}$} orbifolds'' {\em JHEP} {\bf 11} (2000) 022
\href{http://www.arXiv.org/abs/hep-th/0009251}{[{\tt hep-th/0009251}]}.

\bibitem{Conrad:2000cw}
J.~O. Conrad ``On fractional instanton numbers in six dimensional heterotic
  {$\E{8} \times \E{8}$} orbifolds'' {\em Fortsch. Phys.} {\bf 49} (2001)
  455--458
\href{http://www.arXiv.org/abs/hep-th/0101023}{[{\tt hep-th/0101023}]}.

\bibitem{gsw_2}
M.~B. Green, J.~H. Schwarz, and E.~Witten {\em Superstring theory vol. 2:
  {L}oop amplitudes, anomalies and phenomenology}.
\newblock Cambridge, Uk: Univ. Pr. 596 P. (Cambridge Monographs On Mathematical
  Physics) 1987.

\bibitem{Erler:1993zy}
J.~Erler ``Anomaly cancellation in six-dimensions'' {\em J. Math. Phys.} {\bf
  35} (1994) 1819--1833
\href{http://www.arXiv.org/abs/hep-th/9304104}{[{\tt hep-th/9304104}]}.

\bibitem{Berkooz:1996iz}
M.~Berkooz {\em et al.} ``Anomalies, dualities, and topology of {D=6 N=1}
  superstring vacua'' {\em Nucl. Phys.} {\bf B475} (1996) 115--148
\href{http://www.arXiv.org/abs/hep-th/9605184}{[{\tt hep-th/9605184}]}.

\bibitem{Strominger:1986uh}
A.~Strominger ``Superstrings with torsion'' {\em Nucl. Phys.} {\bf B274} (1986)
253.

\bibitem{LopesCardoso:2002hd}
G.~Lopes~Cardoso {\em et al.} ``{Non-Kaehler} string backgrounds and their five
  torsion classes'' {\em Nucl. Phys.} {\bf B652} (2003) 5--34
\href{http://www.arXiv.org/abs/hep-th/0211118}{[{\tt hep-th/0211118}]}.

\bibitem{ibanez_87}
L.~E. Ibanez, H.~P. Nilles, and F.~Quevedo ``Orbifolds and {W}ilson lines''
  {\em Phys. Lett.} {\bf B187} (1987)
25--32.

\bibitem{ibanez_88}
L.~E. Ibanez, J.~Mas, H.~P. Nilles, and F.~Quevedo ``Heterotic strings in
  symmetric and asymmetric orbifold backgrounds'' {\em Nucl. Phys.} {\bf B301}
  (1988)
157.

\bibitem{Polchinski:1995df}
J.~Polchinski and E.~Witten ``Evidence for heterotic - type {I} string
  duality'' {\em Nucl. Phys.} {\bf B460} (1996) 525--540
\href{http://www.arXiv.org/abs/hep-th/9510169}{[{\tt hep-th/9510169}]}.

\bibitem{Kakushadze:1997wx}
Z.~Kakushadze ``Aspects of {N = 1} type {I-}heterotic duality in four
  dimensions'' {\em Nucl. Phys.} {\bf B512} (1998) 221--236
\href{http://www.arXiv.org/abs/hep-th/9704059}{[{\tt hep-th/9704059}]}.

\bibitem{Kakushadze:1998cd}
Z.~Kakushadze, G.~Shiu, and S.~H.~H. Tye ``Type {IIB} orientifolds, {F}-theory,
  type {I} strings on orbifolds and type {I} heterotic duality'' {\em Nucl.
  Phys.} {\bf B533} (1998) 25--87
\href{http://www.arXiv.org/abs/hep-th/9804092}{[{\tt hep-th/9804092}]}.

\bibitem{Angelantonj:1996uy}
C.~Angelantonj, M.~Bianchi, G.~Pradisi, A.~Sagnotti, and Y.~S. Stanev ``Chiral
  asymmetry in four-dimensional open- string vacua'' {\em Phys. Lett.} {\bf
  B385} (1996) 96--102
\href{http://www.arXiv.org/abs/hep-th/9606169}{[{\tt hep-th/9606169}]}.

\bibitem{Aldazabal:1997wi}
G.~Aldazabal, A.~Font, L.~E. Ibanez, A.~M. Uranga, and G.~Violero
  ``Non-perturbative heterotic {D = 6,4, N = 1} orbifold vacua'' {\em Nucl.
  Phys.} {\bf B519} (1998) 239--281
\href{http://www.arXiv.org/abs/hep-th/9706158}{[{\tt hep-th/9706158}]}.

\bibitem{Lalak:1999bk}
Z.~Lalak, S.~Lavignac, and H.~P. Nilles ``String dualities in the presence of
  anomalous {U(1)} symmetries'' {\em Nucl. Phys.} {\bf B559} (1999) 48--70
\href{http://arXiv.org/abs/hep-th/9903160}{[{\tt hep-th/9903160}]}.

\bibitem{Ganor:2002ae}
O.~J. Ganor and J.~Sonnenschein ``On the strong coupling dynamics of heterotic
  string theory on {$C^3/Z(3)$}'' {\em JHEP} {\bf 05} (2002) 018
\href{http://www.arXiv.org/abs/hep-th/0202206}{[{\tt hep-th/0202206}]}.

\end{thebibliography}\endgroup
}

\end{document}